\documentclass[final,3p,times]{elsarticle}
\usepackage{graphicx}
\usepackage{pifont,latexsym,ifthen,rotating,calc,textcase,booktabs,color}
\usepackage{amssymb,amsfonts,amsbsy,amsmath}
\usepackage{mathabx}
\usepackage{float}
\usepackage{upgreek}
\usepackage{subfigure}
\usepackage{textcomp}
\usepackage{placeins}
\usepackage{ifthen}
\usepackage{bigstrut}
\usepackage{lineno}

\def\path{\string~/}
\usepackage{bm}
\usepackage{pgfplots}
\usepackage{tikz}
\usetikzlibrary{shapes, shadows, arrows, positioning,patterns,decorations.pathreplacing}

\journal{Chemical Engineering Science}

\biboptions{sort&compress}

\begin{document}

\begin{frontmatter}

\title{Simulation of PMMA powder flow electrification using a new charging model based on single-particle experiments}

\author[label1,label2]{Holger Grosshans\corref{cor1}}
\ead{holger.grosshans@ptb.de}
\author[label1]{Wenchao Xu}
\author[label3]{Tatsushi Matsuyama}

\cortext[cor1]{Corresponding author}

\address[label1]{Physikalisch-Technische Bundesanstalt (PTB), Braunschweig, Germany}
\address[label2]{Otto von Guericke University of Magdeburg, Institute of Apparatus- and Environmental Technology, Magdeburg, Germany}
\address[label3]{Soka University, Faculty of Engineering, Hachioji, Tokyo, Japan}

\begin{abstract}
Thus far, simulations have failed to predict accurately electrostatic powder charging during pneumatic transport.
We advanced the modeling of powder flow charging by a three-part study:
first, we shot individual particles on a metal target and measured the exchanged charge.
Second, based on these results, we formulated an empirical model and implemented it in our CFD tool.
Using this tool, we performed large-eddy simulations of the powder flow through a square duct with a cylindrical obstacle inside.
Finally, we compared the simulations to measurements in our pneumatic conveying test rig.
The simulations successfully predicted the charging of powder consisting of monodisperse particles of a size of 200~$\upmu$m.
Contrary to the usual procedure for this type of simulation, the tool requires no tuning of any parameters.
According to our simulations, the powder mostly charged when hitting the cylindrical obstacle.
The contacts led to bipolar charge distributions.
\end{abstract}

\begin{keyword}
Simulation, Experiments, Powder flow, Triboelectricity, Bipolar charging
\end{keyword}

\end{frontmatter}


\section{Introduction}

During pneumatic conveying, out of all industrial operations, powder charges by far the strongest~\citep{Kli18}.
If the electrostatic charge on the particles becomes too high, it can discharge and cause a dust explosion.
Dust explosions often destroy complete plants, resulting in severe economic damage and even loss of lives~\citep{Glor03,Osh11,OshaUS}.
Knowing how, where, and why powder electrifies can prevent these explosions.

More precisely, process safety requires finding the relation between the flow conditions, the powder and duct material, and the resulting particle charge.
In an experiment, it is to date impossible to observe non-intrusively the trajectories of individual particles and simultaneously the charge they carry.
Instead, a Faraday measures the total charge of the powder when leaving the pipe~\citep{Wata06,Nda11,Pel18,Gro17b}.
\textcolor{black}{Or, the charge is derived by integrating the current of the conveying line if it is conductive~\citep{Tagh19,Tagh20}}.
Even though these experiments progressed knowledge tremendously, the above-named detailed questions remain unanswered.

These difficulties motivated computing powder electrification by numerical simulations.
Most of these simulations solve the particles in the Lagrangian framework and the carrier gas flow through Reynolds-averaged Navier-Stokes equations~\citep{Kol89}, large-eddy simulations~\citep{Kor14,Gro16a}, or direct numerical simulations~\citep{Gro17a}.
Recently, Eulerian formulations for charging particulates appeared~\citep{Kol18,Ray18,Gro20h}, which reduce the computational expense for systems comprising abundant particles.

A decisive part of these simulations is the model that predicts the charge exchange between one particle and a solid surface upon contact (see the recent review of \citet{Chow21}).
A wide-spread model for that purpose is the so-called condenser model, which initially described the charge exchange between two colliding solid spheres~\citep{Soo71} and later evolved to the impact of a spherical particle on a plane surface~\citep{John80}.
The name refers to its mathematical analogy to the temporal response of a capacitor (\textcolor{black}{formerly} known as condenser) in a resistor-capacitor (R-C) circuit.
Whereas the original formulation assumes a uniform charge distribution on the particle's surface, an extension accounts for the charge distribution on non-conductive particles~\citep{Gro16f}.
Even though the condenser model appears in different versions in the literature, all versions aim to predict the electron transfer driven by the contact potential difference.
However, since it is not clear if this mechanism really controls the charge exchange, and due to the uncertainty of many material parameters, the condenser model usually requires heavy tuning.
Especially, the condenser model fails to predict bipolar charging, which frequently occurs during the flow of polymer particles~\citep{Zhao03,Bil14}.

An alternative to the condenser model is the charge relaxation model~\citep{Mat95b}, wherein the evolution of the potential difference between the particle and the surface determines the charge transfer.
According to this model, the charge uptake of a particle is limited by an immediate discharge.
The discharge takes place at the contact gap when the potential difference exceeds the gaseous breakdown limit.
Another promising approach is computing the charge transfer between two surfaces from first principles~\citep{Shen16}.
However, the laborious explicit modeling of the molecular structure of the involved materials restricts this approach to simple systems of well-ordered structures, such as single-crystal alumina and silicon oxide.
Expanding the approach to polymers is not feasible to date since polymers are structurally heterogeneous, often semi-crystalline, in non-equilibrium states, and contain impurities.

Generally speaking, all the above-discussed models approximate some physical mechanisms while neglecting others.
On the other hand, the exact nature of triboelectric charging is an ongoing debate~\citep{Lacks19}.
For example, it is unclear if ions, electrons, or nanoscale material patches dominate charge transfer between surfaces~\citep{Lacks12}.
Furthermore, the sheer number of parameters affecting contact electrification, including surface roughness~\citep{Ire08}, \textcolor{black}{contact mode~\citep{Ire12}}, geometry~\citep{Qin16}, and humidity~\citep{Nom03}, complicates its theoretical description.
Due to these issues, simulations employing any of the discussed models usually differ from measurements by a large amount or require careful tuning.

The understanding of particle charging advanced significantly through single-particle experiments.
In these experiments, a particle impacts a target with a controlled velocity and angle.
If the impact conditions are constant, and the pre-charge is distributed uniformly on the particle's surface, the impact charge depends linearly on the pre-charge~\citep{Mas83,Yam86}.
On the contrary, the impact charge of insulating particles scatters and is hardly reproducible~\citep{Mat03}.
Partly, it scatters because the charge distributes non-uniformly on the surface of insulating particles.
\textcolor{black}{For the material pair of polymer particles and a metal target, the impact charge is even sometimes positive and sometimes negative~\citep{Mat03}
The seemingly erratic charging of polymer particles hinders their description by a deterministic model.}

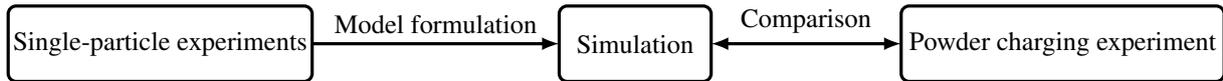
\begin{figure}[t]
\begin{center}
\begin{tikzpicture}[very thick]
\draw [rounded corners=3pt] (0,0) rectangle (4,1) node[midway] {Single-particle experiments};
\draw [->,>=latex] (4,0.5) -- (7.25,.5) node[midway,above]{Model formulation};
\draw [rounded corners=3pt] (7.25,0) rectangle (9.25,1) node[midway] {Simulation};
\draw [<->,>=latex] (9.25,0.5) -- (11.75,.5) node[midway,above]{Comparison};
\draw [rounded corners=3pt] (11.75,0) rectangle (16.0,1) node[midway] {Powder charging experiment};
\end{tikzpicture}
\end{center}
\caption{The idea of the research in this paper:
from the results of single-particle experiments we formulated a model and implemented it in our simulation tool.
The simulations were compared to the measurements in a powder charging experiment.}
\label{fig:idea}
\end{figure}

Figure~\ref{fig:idea} summarizes the idea of the research reported in this paper:
we formulated a model for the charging of PMMA particles.
Opposite to other approaches, we do not aim to quantify physical mechanisms but propose a purely empirical model generated from the results of single-particle experiments.
Then, we implemented this model in our CFD simulation tool and computed the charging of a complete powder in a duct flow.
Finally, we measured the charging of powder flows using our corresponding test rig and compared the results to the simulations.

\section{Methods}

\color{black}
\subsection{Constraints to the size of the particles and duct}

We use two experimental test rigs and one simulation tool.
The overarching aim of the experimental and numerical setups is their consistency.
Each of the three research methods imposes constraints to the size of the particles and duct;
thus, our investigation conditions are defined by the intersection of the optimum ranges of these methods.

More specifically, in the single-particle experiment (Sec.~\ref{sec:single}), to pick up individual particles with tweezers, they need to be at least 100~$\upmu$m in size.
At the same time, to fit through the orifice of the box, they should not be much larger than 200~$\upmu$m.
In the CFD simulations (Sec.~\ref{sec:pafix}), the cross-section of the duct needs to be resolved by a sufficient number of computational cells.
But the Lagrangian approach assumes a negligible particle to cell volume ratio.
In other words, the duct width has to be large compared to the particle.
Finally, to reach a developed flow in the pneumatic conveying experiments (Sec.~\ref{sec:rig}), the duct's length to width ratio must be large.
Meanwhile the laboratory space limits the duct's length, which, therefore, imposes an upper limit to its width.

To fulfill these constraints for 100~$\upmu$m and 200~$\upmu$m particles, we chose a duct width of 45~mm and a length of~2.3~m.
\color{black}

\subsection{Single-particle charging experiment}
\label{sec:single}

\begin{figure}[b]
\centering
\subfigure[]{\includegraphics[trim=0cm 0cm 0cm 40mm,clip=true,width=.4\textwidth]{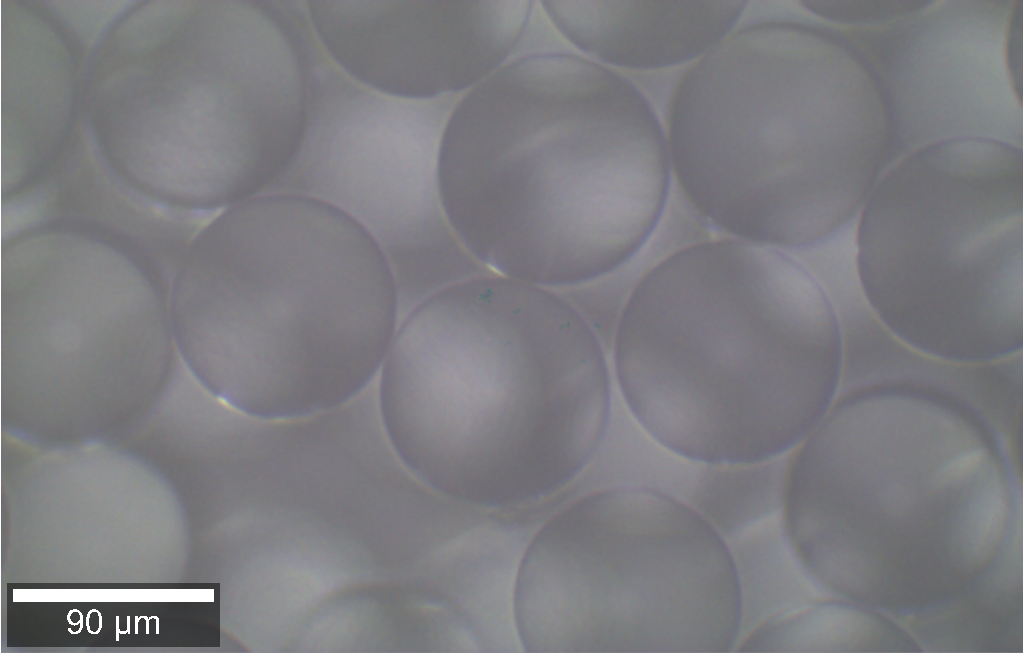}\label{fig:particles}}
\qquad\qquad
\subfigure[]{\includegraphics[trim=0mm 0mm 0mm 0mm,clip=true,width=0.245\textwidth]{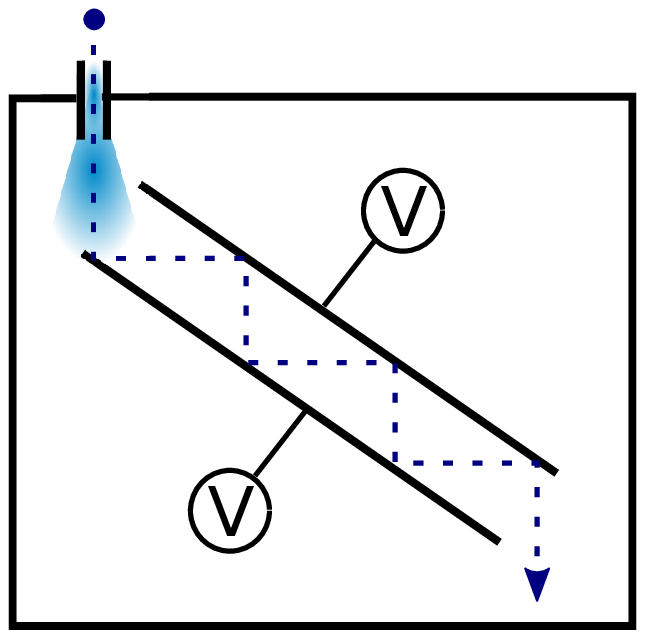}\label{fig:cascade}}
\qquad\qquad
\caption{(a) Spherical PMMA particles of a size of 100~$\upmu$m (reprinted from Ref.~\citep{Gro20c});
(b) schematics of the single-particle experiment.}
\label{fig:}
\end{figure}

\begin{figure}[tb]
\centering
\subfigure[]{\includegraphics[trim=0mm 0mm 0mm 0mm,clip=true,width=0.47\textwidth]{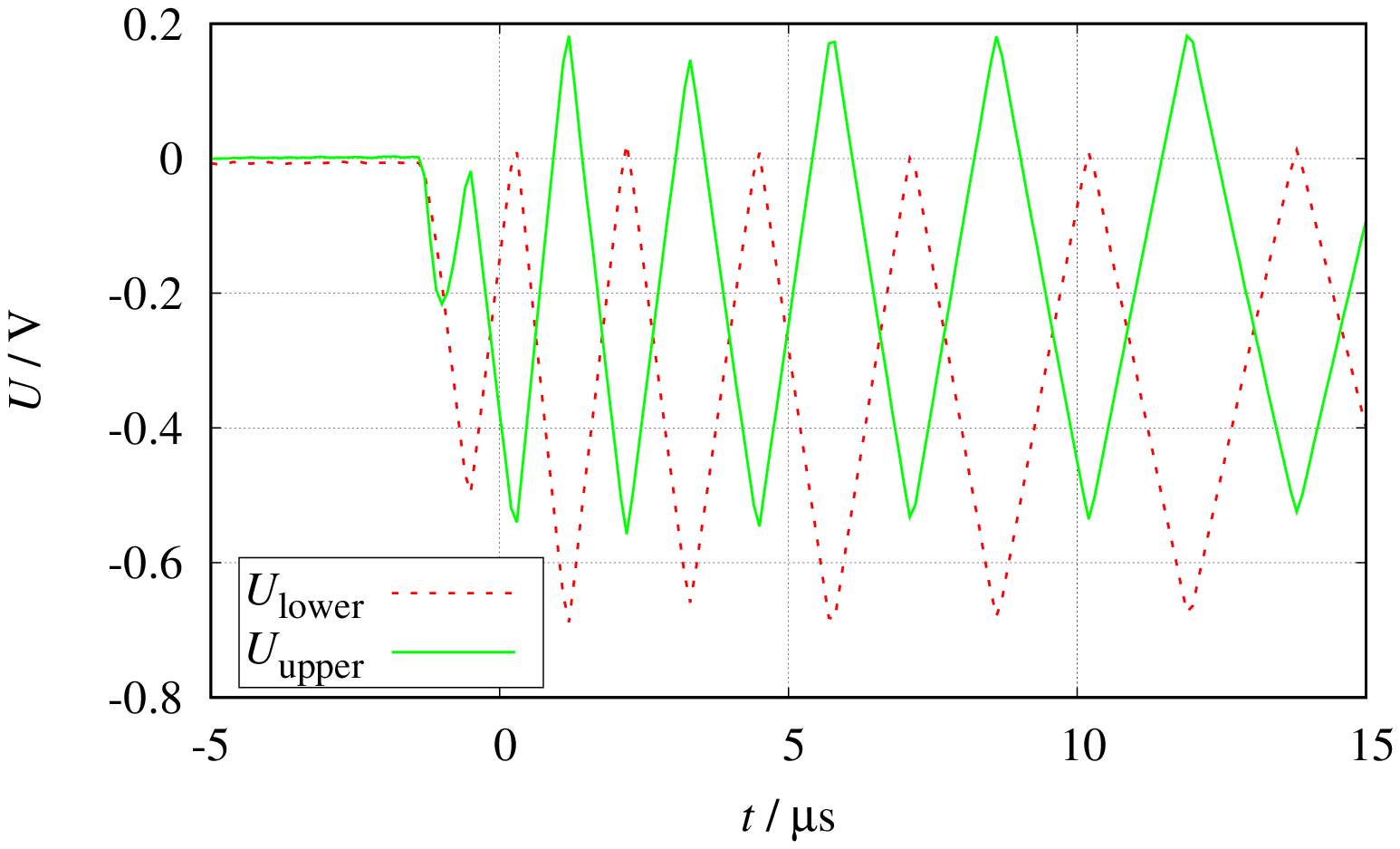}\label{fig:osci}}\quad
\subfigure[]{\includegraphics[trim=0mm 0mm 0mm 0mm,clip=true,width=0.47\textwidth]{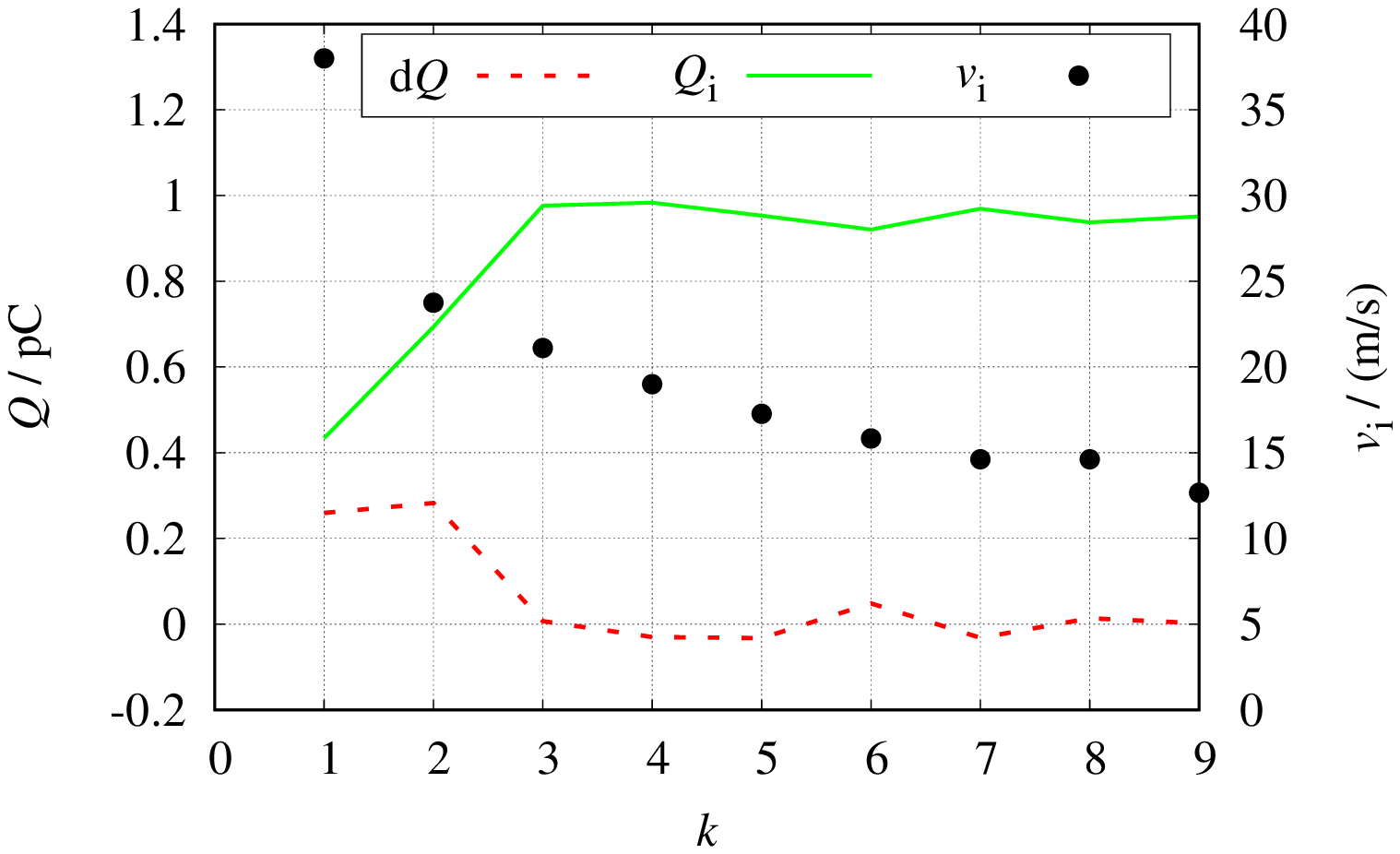}\label{fig:post}}
\caption{(a) Oscilloscope readings and (b) impact data after post-processing of one experiment with a 200~$\upmu$m PMMA particle.}
\label{fig:}
\end{figure}

To develop the particle charging model, we measured the charging of individual particles upon impact on a target.
The particles were of a size of 100~$\upmu$m, see Fig.~\ref{fig:particles}, or 200~$\upmu$m, spherical, and made of PMMA.
\textcolor{black}{
The model aims to predict the charging of the particles in the flow experiments (Sec.~\ref{sec:rig}).
Thus, the flow experiments convey particles of identical powder batches as the single-particle experiments.
Further, to ensure consistent surface conditions, both experiments used particles as delivered, which means without any further treatment.}

Figure~\ref{fig:cascade} depicts the schematics of the single-particle experiment.
This apparatus was developed and described by \citet{Mur13} and is outlined in the following.
For each test, one particle is picked with tweezers and dropped into the orifice at the top.
Moving air strongly disturbs such light particles' trajectories.
\textcolor{black}{
To control the trajectories, an acrylic box encloses the apparatus, and a vacuum pump evacuates the air from the inside.
The pressure between the inside and outside of the box differs by up to 0.85 atmospheres.
Due to the pressure difference, an air jet forms at the box orifice, which accelerates the particle toward the target.}

The target consists of two parallel \textcolor{black}{steel} plates mounted at a distance of $h=19$~mm to each other and an angle of 45$^{\circ}$ to the velocity vector of the incoming particle.
Because of this design, the particle bounces between the upper and lower plates, as sketched in Fig.~\ref{fig:cascade}.
Two independent charge amplifiers connect each of the two plates to an oscilloscope.

Figure~\ref{fig:osci} displays typical oscilloscope recordings of one impact experiment after smoothing the signals.
For smoothing, a box filter of the width of 1~ns was applied.
From these data, we computed the charge of the particle before each impact ($Q_\mathrm{i}$), the impact charge (d$Q$), and the impact velocity ($v_\mathrm{i}$), see Fig.~\ref{fig:post}.
The first step of the post-processing is to convert the voltage time-histories recorded at both plates (Fig.~\ref{fig:osci}) into charge time-histories ($Q_\mathrm{upper}$ and $Q_\mathrm{lower}$) by multiplying them with the capacity of the system.
This charge is the sum of the charge transferred to the plate during an impact, i.e., -d$Q$, and the mirror charge induced by the particle.

The peaks of the curves in Fig.~\ref{fig:osci} indicate the particle impacts on the plates.
The charge held by the particles before the $k$-th impact follows from the amplitude of these peaks by
\begin{equation}
\label{eq:Qi}
Q_\mathrm{i}(k)=
\left\{ 
\begin{matrix}
Q_\mathrm{upper}(k)-Q_\mathrm{upper}(k-1) & \text{for} \quad k={2,4,6, \ldots} \\
Q_\mathrm{lower}(k)-Q_\mathrm{lower}(k-1) & \text{for} \quad k={1,3,5, \ldots}
\end{matrix}
\right.
\end{equation}
According to the above equations, the calculation of $Q_\mathrm{i}(1)$ requires the knowledge of $Q_\mathrm{lower}(0)$.
We estimated $Q_\mathrm{lower}(0)$ from the first minimal point of the upper plate's curve, in Fig.~\ref{fig:osci} at about -1~$\upmu$s.
Contrary to all other peaks, no simultaneous peak appears at the lower plate.
The peak at the lower plate is missing because $k=0$ is not an actual impact but the closest approximation of the particle to the upper plate on its fly-by from the inlet of the box to the lower plate (cf.~Fig.~\ref{fig:cascade}).

The impact charge follows directly from the results of Eq.~(\ref{eq:Qi}),
\begin{equation}
\label{eq:dQk}
\mathrm{d}Q(k)= Q_\mathrm{i}(k) - Q_\mathrm{i}(k-1).
\end{equation}
Finally, the ratio of the distance between the plates and the time delay between two successive impacts approximates the impact velocity, 
\begin{equation}
\label{eq:vi}
v_\mathrm{i}(k)= h/\left(t_\mathrm{i}(k) - t_\mathrm{i}(k-1)\right).
\end{equation}
This approximation neglects the deceleration of the particle between two successive impacts, which is reasonable due to the low pressure in the box.

The post-processed data in Fig.~\ref{fig:post} reveal that the particle flies the fastest before the first impact and slows down while moving through the apparatus.
The impact charge is high during the first and second impact, whereas nearly no charge transfers to the plates afterward.
The reason for that behavior becomes clear from the $Q_\mathrm{i}$ curve:
the particle charge after the second impact, which is about 1~pC, is apparently close to its saturation charge.
Therefore, it takes up or dispenses only a little charge in following contacts with the plates.

\subsection{Powder flow electrification test rig}
\label{sec:rig}

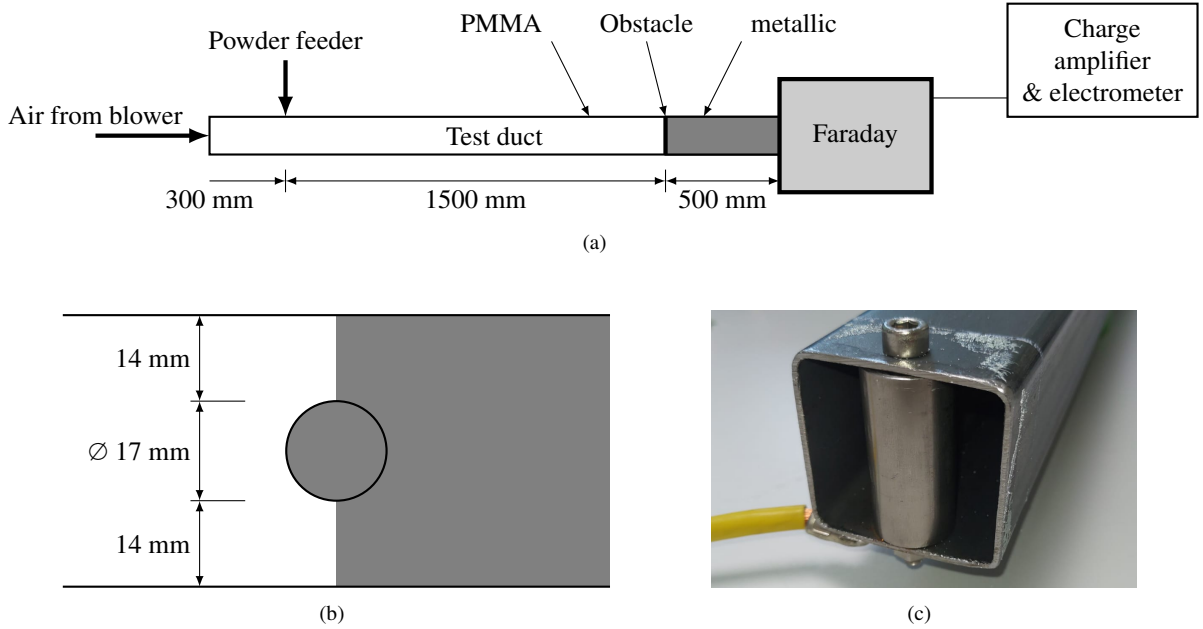
\begin{figure}[tb]
\centering
\subfigure[]{
\begin{tikzpicture}[thick]
\draw [->,>=latex,ultra thick] (0,0.25) node[right,above]{Air from blower} -- (1.5,.25);
\draw [] (1.5,.0) rectangle (9,.5) node[midway] {Test duct};
\draw [fill=gray] (7.5,.0) rectangle (9,.5) ;
\draw [ultra thick] (7.5,0) -- (7.5,.5);
\draw [->,>=latex,thin] (6,1.5) node[above left,align=right]{PMMA} -- (6.5,.5);
\draw [->,>=latex,thin] (7.25,1.5) node[above]{Obstacle} -- (7.5,.5);
\draw [->,>=latex,thin] (8.5,1.5) node[above right,align=left]{metallic} -- (8.0,.5);
\draw [->,>=latex,ultra thick] (2.5,1.25) node[right,above]{Powder feeder} -- (2.5,.5);
\draw [fill=gray!40,ultra thick] (9,-.5) rectangle (11,1) node[midway] {Faraday};
\draw [thin] (11,.75) -- (12,.75);
\draw [] (12,0.5) rectangle (14.5,2) node[align=center,midway] {Charge\\amplifier\\\& electrometer};
\draw [->,>=latex,thin] (1.5,-.35) node[below,align=right] {300 mm} -- (2.5,-.35);
\draw [thin] (2.5,-.2) -- (2.5,-.5);
\draw [<->,>=latex,thin] (2.5,-.35) -- (7.5,-.35) node[below,midway,align=right] {1500 mm};
\draw [thin] (7.5,-.2) -- (7.5,-.5);
\draw [<->,>=latex,thin] (7.5,-.35) -- (9,-.35) node[below,midway,align=right] {500 mm};
\end{tikzpicture}
\label{fig:rig}
}\\ 

\bigskip
\subfigure[]{
\begin{tikzpicture}[thick,scale=1.2]
\path [fill=gray] (4,0) rectangle (7,3);
\draw [fill=gray] (4,1.5) circle [radius=.55];
\draw [] (1,0) -- (7,0);
\draw [] (1,3) -- (7,3);
\draw [<->,>=latex,thin] (2.5,3) -- (2.5,2.05) node[left,midway,align=right] {14 mm};
\draw [thin] (2.4,2.05) -- (3.,2.05);
\draw [<->,>=latex,thin] (2.5,2.05) -- (2.5,0.95) node[left,midway,align=right] {$\diameter$ 17 mm};
\draw [thin] (2.4,0.95) -- (3.,0.95);
\draw [<->,>=latex,thin] (2.5,0) -- (2.5,0.95) node[left,midway,align=right] {14 mm};
\end{tikzpicture}
\label{fig:obstacle}
}
\qquad
\quad
\subfigure[]{
\includegraphics[trim=85mm 120mm 125mm 70mm,clip=true,width=0.34\textwidth]{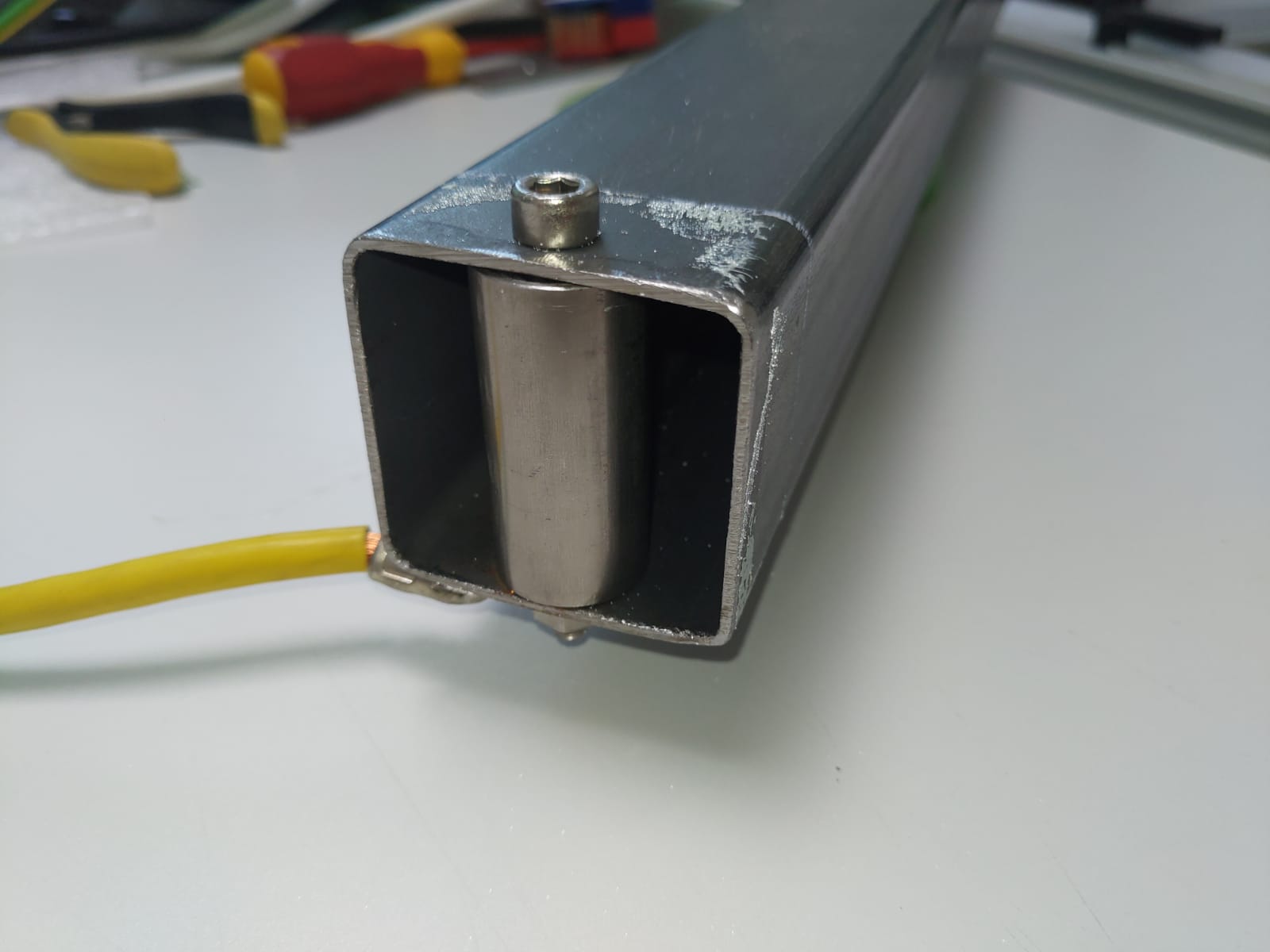}
\label{fig:photo}
}
\caption{Schemata of the powder flow electrification test rig.
(a) Side view:
the air and powder flows horizontally from left to right through the duct .
The Faraday measures the charge accumulated by the powder.
(b) Top view:
enlargement of the geometry of the cylindrical obstacle.
The part of the duct upstream of the obstacle is made of PMMA, the downstream part is metallic.
(c) Photo of the obstacle installed in the metallic part of the duct.
}
\label{fig:powelec}
\end{figure}

We measured the electrification of powder flows in a previously developed test rig~\citep{Gro20c}.
We slightly modified the test rig schematically shown in Fig.~\ref{fig:rig} for the current study from the original set-up.
The blower presses ambient air into the horizontal test duct.
The test duct is on its complete length of a square-shaped cross-section of an inner side length of 45~mm.
After a short development section, the vibrational feeder seeds the powder at a constant rate into the airflow.

The carrier flow conveys the powder through the duct's 1500~mm long PMMA part.
Afterward, the powder flow enters the 500~mm long \textcolor{black}{steel} part.
At the beginning of this \textcolor{black}{steel} part, a vertically oriented cylindrical obstacle is installed, also \textcolor{black}{made of steel.
Thus, the material pair when a particle contacts the metallic part is the same as in the single-particle experiment, namely PMMA and steel.}
As shown in Fig.~\ref{fig:obstacle}, the obstacle blocks approximately one-third of the total cross-sectional area of the duct.
Figure~\ref{fig:photo} shows a photo of the obstacle and the beginning of the metal part.
The yellow cable in the bottom left of the image connects the duct and the obstacle to the ground.

At the end of the duct, a filter located inside the Faraday separates the powder from the air stream.
The charge amplifier connects the Faraday to the electrometer, which displays the charge inside the Faraday.
The charge leakage time-scale of the electrical circuit is small compared to the time-scale of the charge increase during conveying.
A detailed specification of the equipment is given in Ref.~\citep{Gro20c}.

As elaborated in the following paragraphs, the experiment is set up and operated as consistently as possible to the simulations.
For this purpose, the PMMA duct is long enough to generate a developed air-particle flow.
A developed flow depends only on physics and not on the blower or the powder feeder, respectively the inflow conditions imposed on the simulations.
Thus, the charging of the powder when passing the obstacle and the metallic duct part, which is the actual topic of our investigation, is independent of the technical details of the facility.
In particular, the flow is comparable between our experiments and simulations.

While being transported for 1500~mm, which means more than thirty times the duct diameter, the turbulent airflow develops.
Also, the particles adopt their preferential positions.
Or at least a position close to that since particle positions can take longer to develop than the airflow.
Therefore, it is not required to model the details of the feeding system in the simulations to reach comparable flow conditions at the beginning of the metal duct.
Also, we start up the air blower before beginning the particle feed to ensure that the start-up does not affect the particle transport.
Then, we feed particles for ten seconds.
By doing so, we ensure that number of particles belonging to the instationary head of the transported powder batch is small compared to the number of particles transported under stationary conditions.

Further, we chose a PMMA duct for the flow development to limit the charge on the particles when entering the metal duct~\citep{Gro20c}.
For the flow conditions of the present paper and the 200~$\upmu$m particles, the absolute specific charge when leaving the PMMA duct was about 5~$\upmu$C/kg to 8~$\upmu$C/kg.
As seen below in the Results section, this charge is small compared to the charge accumulated in the metallic part.
Thus, the charging in the PMMA duct plays a minor role in the total charging and requires no dedicated modeling.
To reiterate, the design decisions and procedures described in this and the previous paragraph aim to maximize the comparability between the experiments and simulations.

\subsection{Powder flow solver}
\label{sec:pafix}

We used our tool \citet{pafiX} to compute the powder flow.
The mathematical model and numerical methods implemented in this tool were exhaustively described by \citet{Gro20b} and are briefly summarized in the following.

The tool solves the Navier-Stokes equations for the carrier gas, Gauss's law for the electric field, and Newton's laws of motion for the dynamics of the particles.
The Navier-Stokes equations, i.e., mass and momentum conservation for incompressible fluids, read
\begin{subequations}
\begin{equation}
\label{eq:mass}
\nabla \cdot {\bm u} = 0
\end{equation}
\begin{equation}
\label{eq:mom}
\frac{\partial {\bm u}}{\partial t} + ({\bm u} \cdot \nabla) {\bm u}
= - \frac{1}{\rho} \nabla p  + \nu \nabla^2 {\bm u} + {\bm F}_{\mathrm s} + {\bm F}_{\mathrm vb} \, .
\end{equation}
\end{subequations}
Herein, $\bm u$ denotes the velocity, $\rho$ the density, $p$ the pressure, and $\nu$ the kinematic viscosity of the fluid. 
The source term ${\bm F}_{\textrm s}$ accounts for the momentum transfer from the particles to the fluid.

Equations~(\ref{eq:mass}) and~(\ref{eq:mom}) were discretized via the finite difference method on a Cartesian grid.
The velocity derivatives in Eq.~(\ref{eq:mass}) and the pressure gradient and viscous terms in Eq.~(\ref{eq:mom}) are discretized via fourth-order central differences, and the convective term in Eq.~(\ref{eq:mom}) by a fifth-order accurate WENO scheme~\citep{Jang96}.
Naturally, nodes of the rectilinear grid do not coincide with the boundary of the cylindrical obstacle.
Thus, we opted for the virtual-boundary method~\cite{Gol93,Rev01} where the solid boundary is replaced by the body force 
\begin{equation}
\label{eq:ib}
{\bm F}_\mathrm{vb}= -C_1 \, {\bm u} \, \mathrm{e}^{-C_2 z^2}
\end{equation}
which is the last term in the above momentum equation.
This source term enforces no-slip boundary conditions at the virtual interface between the solid and the fluid.
We set the constants to $C_1=1/$s and $C_2=6/\Delta h^2$ since this choice leads to satisfying numerical efficiency and stability~\citep{Rev13,Sza16,Gro16h,Gro17d}.
Therein, $h$ is the local cell size in the direction the momentum equation is solved.
Further, $z$ is the distance between the grid node and the surface of the obstacle.
Thus, ${\bm F}_\mathrm{vb}$ becomes negligible at distances larger than two cells from the boundary.

Gauss's law is given by
\begin{equation}
\label{eq:gauss}
\nabla \cdot {\bm E} = \dfrac{\rho_{\mathrm{el}}}{\varepsilon} \, .
\end{equation}
In this equation, $\rho_{\mathrm{el}}$ is the electric charge density, and $\varepsilon$ is the electrical permittivity of free space. 
The derivative in Eq.~\ref{eq:gauss} is discretized via second-order central differences.

The position of the particles is calculated in the Lagrangian frame of reference, i.e., for every particle Newton's second law is solved,
\begin{equation}
\label{eq:newton}
\dfrac{\mathrm{d} \bm u_{\textrm p}}{\mathrm{d} t} = {\bm f}_\mathrm{d} + {\bm f}_\mathrm{coll} + {\bm f}_\mathrm{el}\, .
\end{equation}
Here, ${\bm f}_\mathrm{d}$ is the particle's acceleration by drag, ${\bm f}_\mathrm{coll}$ by collisions with a solid surface or another particle, and ${\bm f}_\mathrm{el}$ by electric field forces.
The last term in Eq.~(\ref{eq:newton}) is calculated as
\begin{equation}
\label{eq:fel}
{\bm f}_\mathrm{el} = \dfrac{Q \, {\bm E}}{m_\mathrm{p}}
\end{equation}
where $Q$ is the charge and $m_\mathrm{p}$ the mass of the particle.
For particles residing in the same computational cell, their electrostatic forces acting on each other computed by Coulomb's law~\citep{Gro17e} are added to ${\bm f}_\mathrm{el}$~\citep{Gro17e}.

The computational domain covers the PMMA duct starting from the feeding point up to the end of the metallic duct (cf.~Fig.~\ref{fig:rig}, thus, a length of two meters.
The mesh consists of 500~cells in the flow direction and 60~cells in both spanwise directions, totaling 1.8~million cells.
Each computation was distributed on ten processors.
This resolution is not fine enough to resolve all flow scales.
Instead, the grid acts as an implicit filter that resolves only the large turbulent eddies, which means we perform large-eddy simulations.
In our simulations, the truncation error of the discretization schemes balances the dissipation of the unresolved small scales.
To this end, we carefully chose the discretization schemes and validated our implicit turbulence modeling method in several studies during the last decade~(e.g.,~\citep{Gro14e,Gro16a}).

The simulations neglect the curvature of the corners of the duct, which is for the metallic part visible in Fig.~\ref{fig:photo}.
As described in Sec.~\ref{sec:rig}, the test rig was designed to simplify the simulations' boundary conditions.
The inlet profile of the carrier gas follows the average velocity of a fully developed turbulent flow superimposed by random velocity fluctuations.
The centreline velocity at the inlet was 14.7~m/s, and the Reynolds number based on the mean velocity of 11.6~m/s was 36\,000.
Even though this profile might be different from the one in the experiment, the discrepancy gets lost during the flow development until the metallic duct part, which is the actual topic of our investigation.
Similarly, the particles are seeded at random positions at the inlet plane but adopt their preferential location during the flow development.

\subsection{Empirical single-particle charging model}
\label{sec:singlem}

We formulated an empirical charging model based on the results of the single-particle experiment described in Sec.~\ref{sec:single}.
The idea of our model is to approximate the impact charge of a particle in the simulation directly from similar impacts measured in the experiment.
That means, before applying our model to another powder, single-particle experiments would have to characterize this specific powder first.
This approach differs from previous particle charging models that try to model physics and require tuning.

To quantify the similarity of a simulated and measured impact, we introduce the non-dimensional distance function
\begin{equation}
\label{eq:dist}
\phi_n= \left[
\left(\dfrac{v_\mathrm{i,sim}-v_{\mathrm{i},n}}{\hat{u}}\right)^2
+ 
\left(\dfrac{Q_\mathrm{i,sim}-Q_{\mathrm{i},n}}{\hat{Q}}\right)^2
\right]^{1/2} \, .
\end{equation}
Herein, $v_\mathrm{i,sim}$ is the simulated particle velocity projected on the surface normal, and $Q_\mathrm{i,sim}$ is its charge right before that impact.
The corresponding quantities of the $n$-th impact in the experiment are $v_{\mathrm{i},n}$ and $Q_{\mathrm{i},n}$.
The distance in both dimensions is normalized by the characteristic scales of the velocity, $\hat{u}$, and charge, $\hat{Q}$.
We define $\hat{u}$ by the average centreline velocity of the gas phase and $\hat{Q}$ by a typical particle charge value.
While the former is an input parameter to the simulations, the latter is not known a-priori.
Based on the results of the single-particle experiments presented in Sec.~\ref{sec:ressingle}, we define $\hat{Q}=1$~pC.

Thus, elements of the vector $\phi$ compare one simulated impact to each of the $n$ measured impacts.
Then, the ten closest experimental impacts, i.e., the elements with the lowest $\phi$ value, are chosen and included in the approximation of the impact charge.
Each of these ten impacts is assigned a weight, $\omega$.
The weight decreases with a Gaussian distribution function depending on its distance to the simulated impact,
\begin{equation}
\label{eq:weight}
\omega_n= \dfrac{\mathrm{e}^{-\phi_n^2}}{\sum_{n=1}^{10} \mathrm{e}^{-\phi_n^2}} \, .
\end{equation}
According to the above expression, the weights are normalized so their sum equals unity.

Finally, the impact charge predicted by the model upon one impact in the simulations is the sum of the products of the ten weights and impact charges, i.e.,
\begin{equation}
\label{eq:dQ}
\mathrm{d}Q_\mathrm{p}= 
\sum\limits_{n=1}^{10}
\omega_n \mathrm{d}Q_n \, .
\end{equation}

To summarize, we blend the outcome of the ten most similar impacts from the single-particle experiments.
Due to the decay of the weights with $\phi$, we observed in the simulations that d$Q_\mathrm{p}$ is dominated by the closest 3--4 impacts.
That means that including more than ten measured impacts would not affect the predicted impact charge.
The variable $\phi$ quantifies the distance of one simulated to the experimental impact conditions in two-dimensional space, one dimension being the impact velocity and the other the particle charge before the impact.
\textcolor{black}{As elaborated above, in reality d$Q_\mathrm{p}$ depends on many more parameters, such as the ambient humidity, temperature, pressure, surface roughness, impact angle, \textcolor{black}{and impact mode.}
Since they cannot be varied in our single-particle experiment, $\phi$ does not reflect these parameters.
However, out of all mentioned parameters, we assume the impact velocity and particle pre-charge to affect the impact charge the strongest.
Therefore, our proposed formulation is the natural first step toward a more general empirical model.}

To reiterate, we model and use the same batch of PMMA particles.
Thus, our model fully covers the influence of the particle material, size, and roughness.

\section{Results \& Discussion}

\textcolor{black}{The first part of this section discusses the single-particle experiments that fill the charging model's database.
The second part presents the CFD simulations using this model and their comparison to powder charging experiments.}

\subsection{Single-particle charging experiments}
\label{sec:ressingle}

\begin{figure}[p]
\centering
\subfigure[$d_\mathrm{p}=100~\upmu$m]{\includegraphics[trim=0mm 0mm 0mm 0mm,clip=true,width=0.47\textwidth]{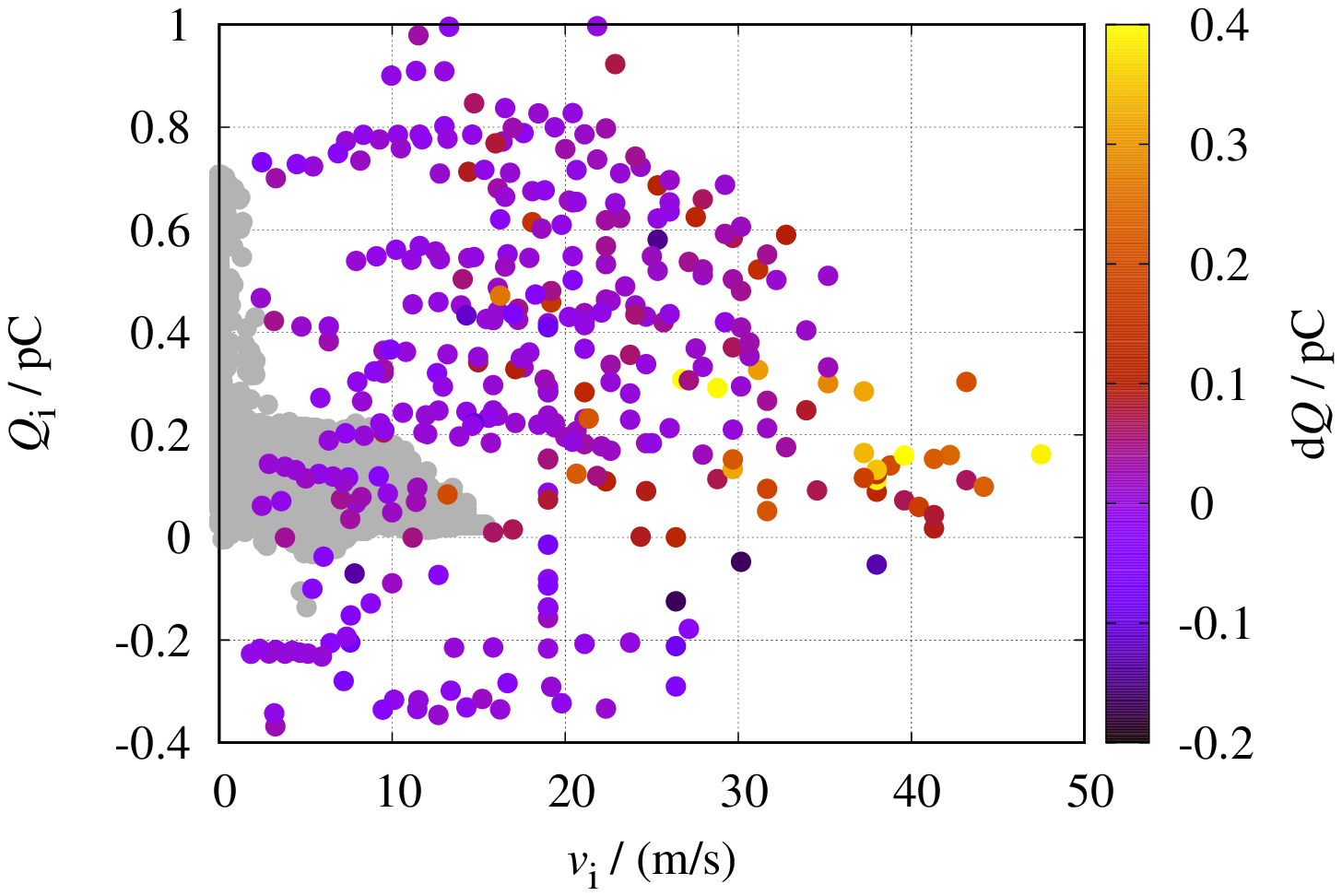}\label{fig:a}}\quad
\subfigure[$d_\mathrm{p}=200~\upmu$m]{\includegraphics[trim=0mm 0mm 0mm 0mm,clip=true,width=0.47\textwidth]{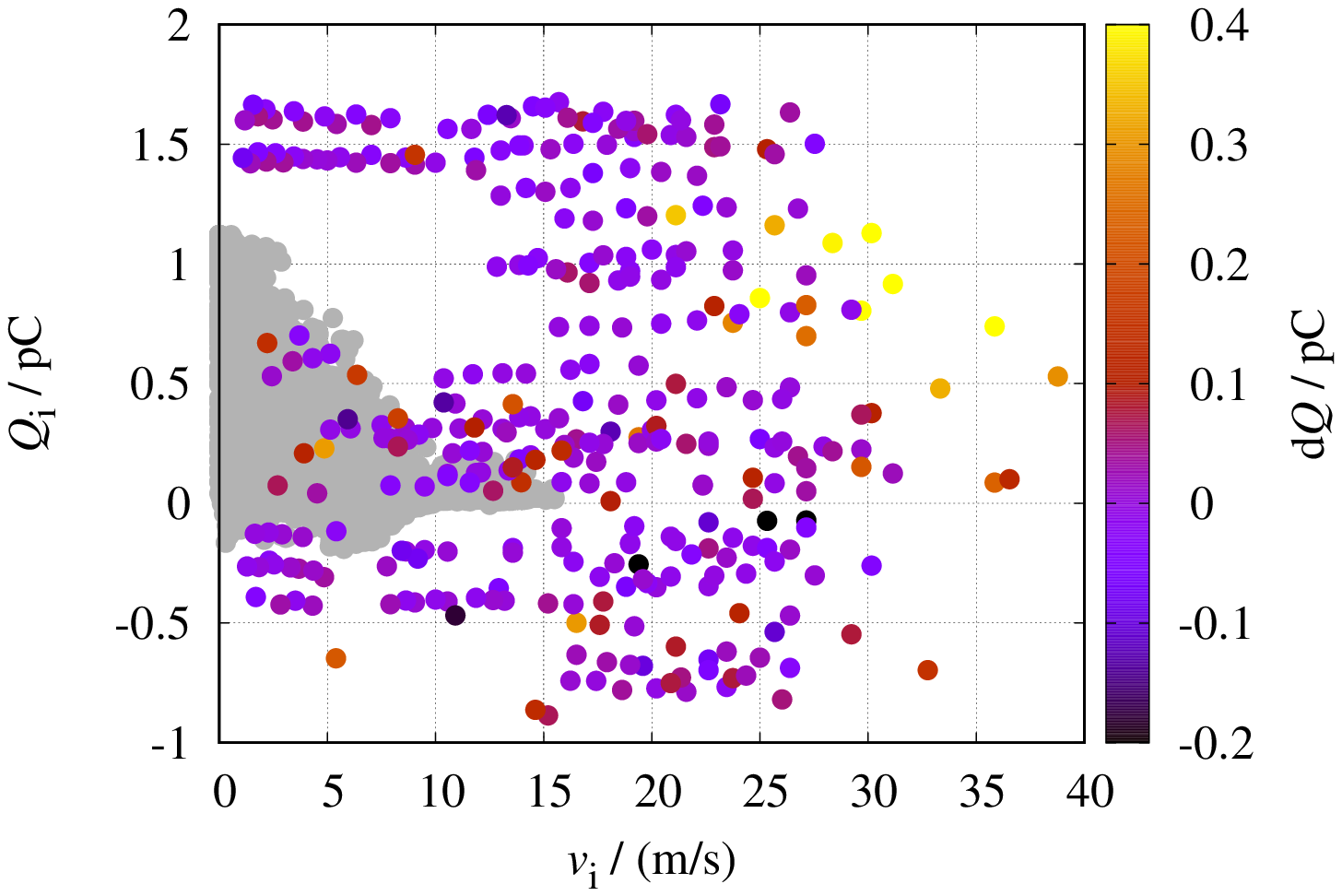}\label{fig:b}}\\
\subfigure[$d_\mathrm{p}=100~\upmu$m]{\includegraphics[trim=0mm 0mm 0mm 0mm,clip=true,width=0.47\textwidth]{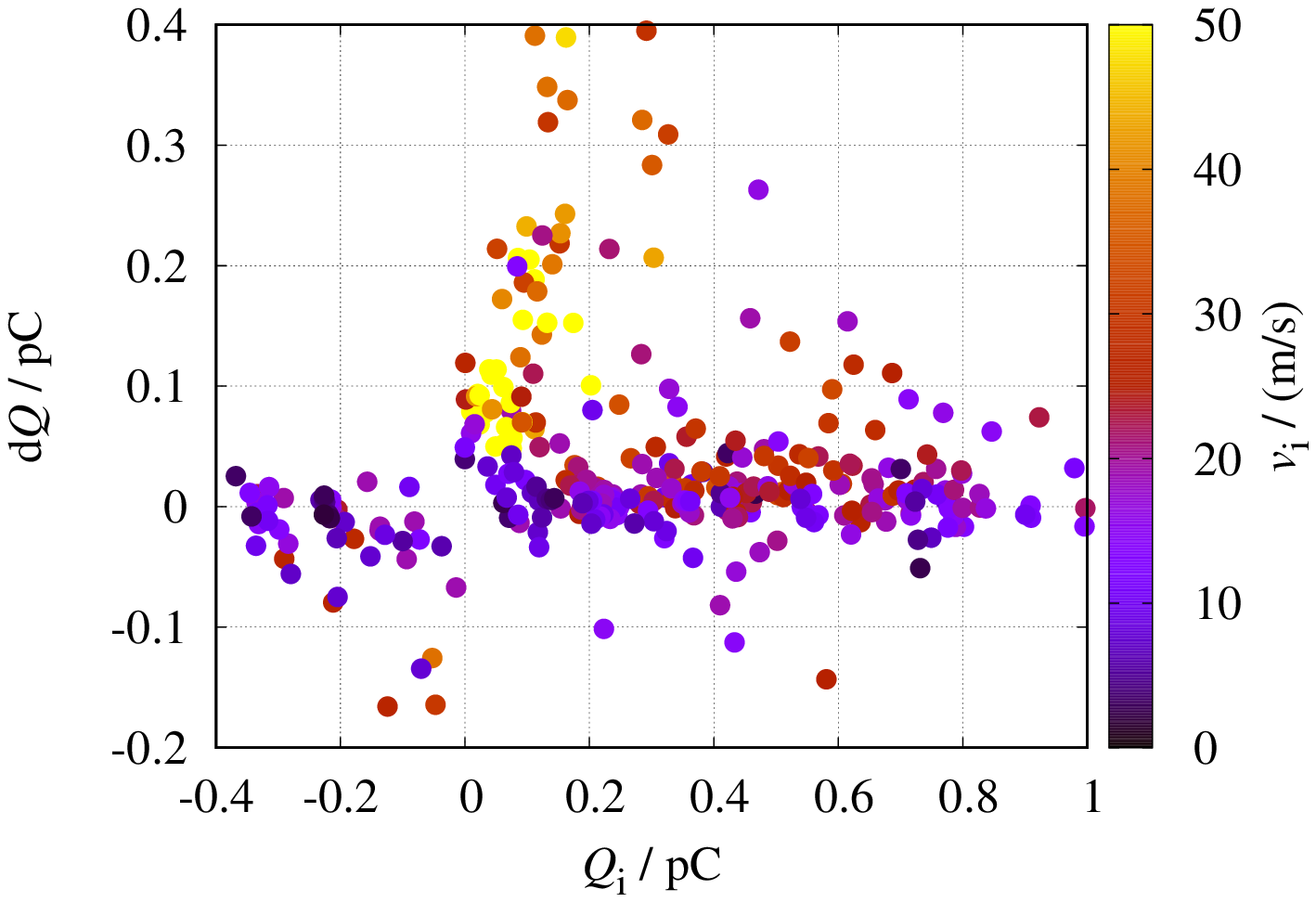}\label{fig:c}}\quad
\subfigure[$d_\mathrm{p}=200~\upmu$m]{\includegraphics[trim=0mm 0mm 0mm 0mm,clip=true,width=0.47\textwidth]{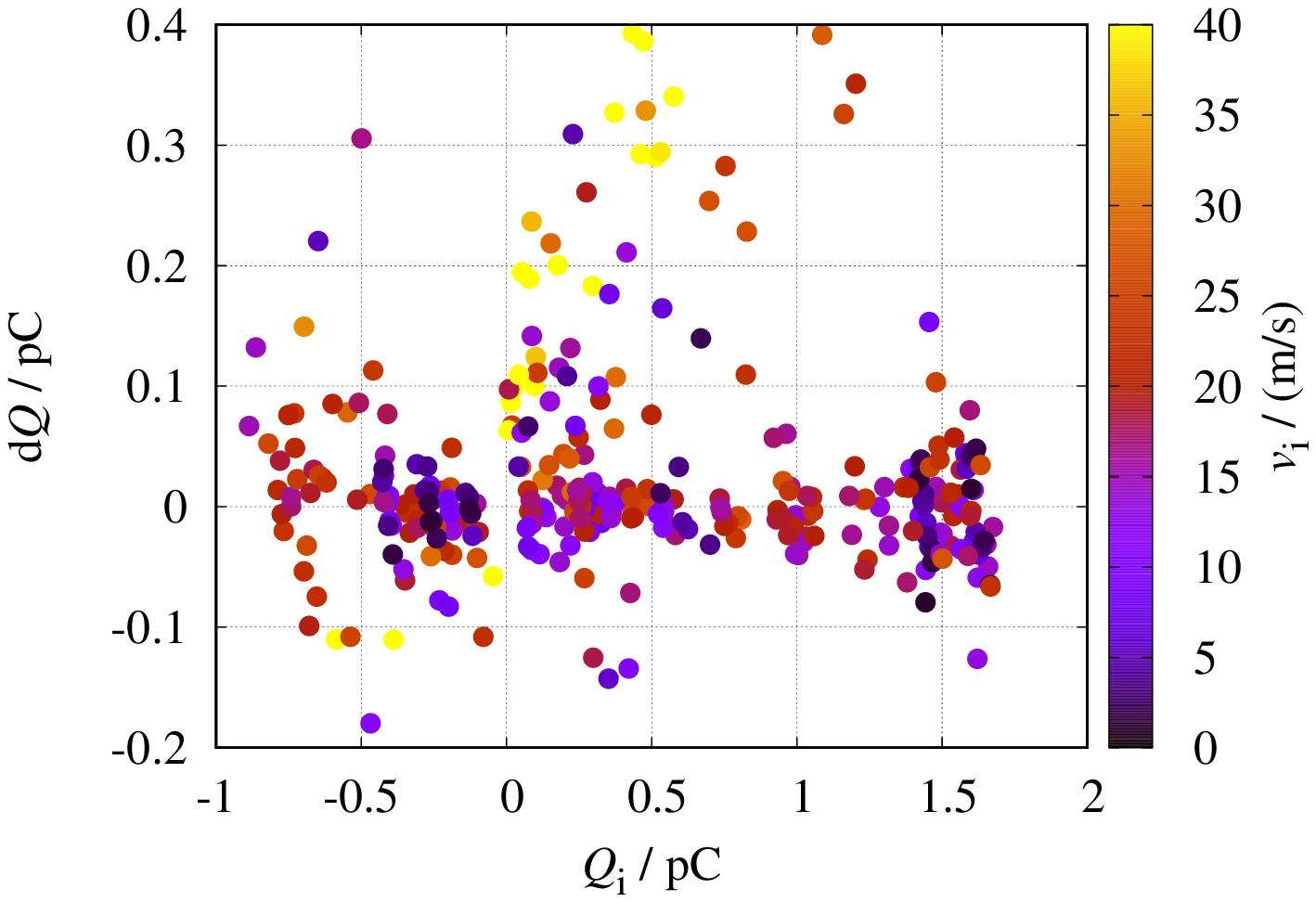}\label{fig:d}}\\
\subfigure[$d_\mathrm{p}=100~\upmu$m]{\includegraphics[trim=0mm 0mm 0mm 0mm,clip=true,width=0.47\textwidth]{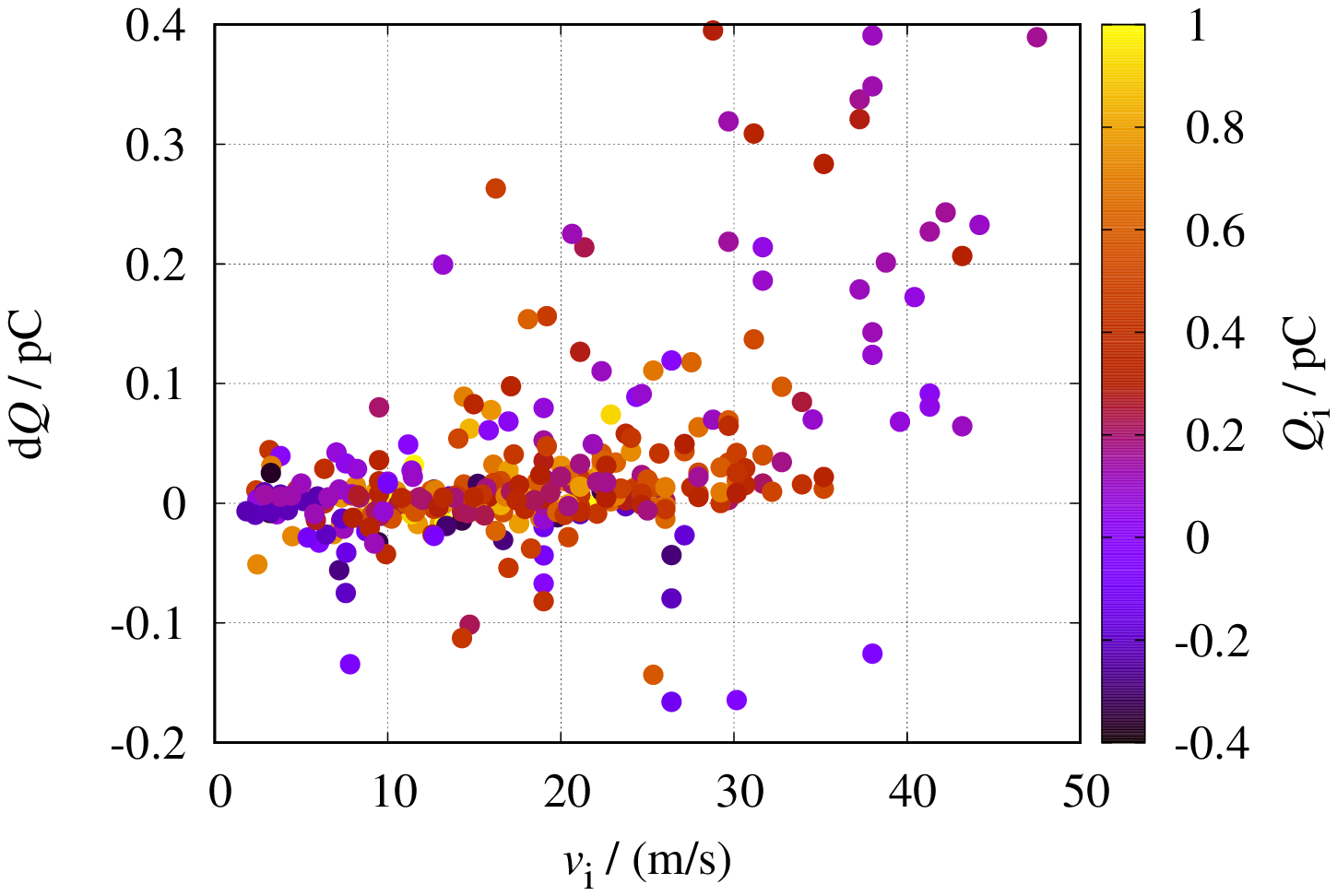}\label{fig:e}}\quad
\subfigure[$d_\mathrm{p}=200~\upmu$m]{\includegraphics[trim=0mm 0mm 0mm 0mm,clip=true,width=0.47\textwidth]{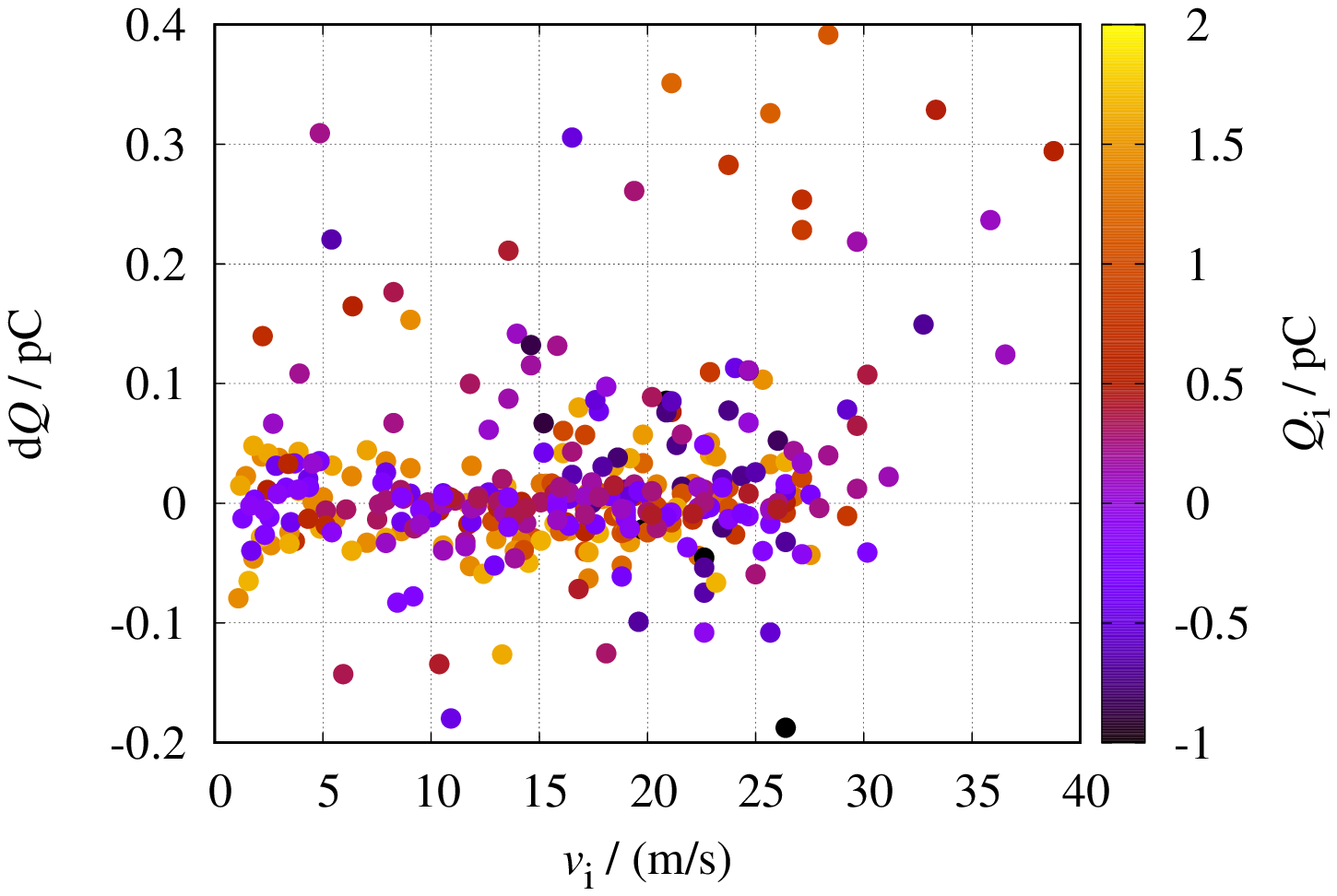}\label{fig:f}}
\caption{Results of the single-particle charging experiment in terms of impact velocity ($v_\mathrm{i}$), initial charge ($Q_\mathrm{i}$), and impact charge (d$Q$).
The left column shows the smaller and the right one the larger particles.
The gray areas in~(a) and~(b) represent the conditions of 15\,000 impacts in the powder flow simulations.
}
\label{fig:ress}
\end{figure}

Single-particle charging experiments delivered the data for the empirical charging model.
We used two different PMMA powder batches for the single-particle experiments, one containing particles of a size of 100~$\upmu$m and the other of 200~$\upmu$m.
With the smaller particles, 38 tests resulted in 358 recorded impacts, and with the larger particles, 56 tests led to 366 impacts.
That means the smaller particles impact on average 9.4~times per experiment and the larger ones only 6.5~times.
\textcolor{black}{
During the tests, the average temperature in the laboratory was 23.5~$^{\circ}$C, varying between 24.5~$^{\circ}$C and 19.9~$^{\circ}$C.
The average relative humidity was 55.3\%, varying between 67.7\% and 38.7\%.
Since the air density in the box was low, these ambiance variations did probably not influence the measurements.}

The left column of Fig.~\ref{fig:ress} shows the impact data of the 100~$\upmu$m and the right column of the 200~$\upmu$m particles.
The three rows of Fig.~\ref{fig:ress} present the same impact data, but the variables are plotted differently against each other.
The axes of the figures are scaled to show approximately 95\% of all impacts.
The 5\% outliers do not influence the model since the distance function $\phi$ in Eq.~(\ref{eq:dist}) becomes large.
Additionally, the gray areas in Figs.~\ref{fig:a} and~\ref{fig:d} outline the range of impact conditions in the CFD simulations, which we discuss in Sec.~\ref{sec:respow}.

\textcolor{black}{
Overall, the impact data Fig.~\ref{fig:ress} appear unstructured.
The lack of obvious structure discourages the formulation of a deterministic charging model.
Instead, it puts forward an empirical model.}

The charge of the particles before the impact is mostly positive, but also negative charges occur.
More precisely, 85\% of the impacts of 100~$\upmu$m and 69\% of the impacts of the 200~$\upmu$m particles take place with a positive pre-charge ($Q_\mathrm{i}>0$).
\textcolor{black}{
But the negative pre-charges do not contradict the triboelectric series, which implies a positive charging of PMMA upon contact with metal.
Some particles are already negatively charged when entering the experiment.
Before entering the experiment, they contact the storage container, other particles of the powder batch, the laboratory table, the tweezers, and repeatedly the inner sides of the orifice.
During these contacts with different materials, the particles collect charge of both polarities;
thus, some particles naturally have a negative pre-charge during the early impacts of the experiment.}

Moreover, Figs.~\ref{fig:e} and~\ref{fig:f} reveal even negative impact charges.
In numbers, for 31\% of the small and 44\% of the large particle impacts, d$Q$ is negative.
\textcolor{black}{These negative impact charges take place after a particle exceeds its equilibrium charge.}
Recalling Fig.~\ref{fig:post}, after reaching equilibrium, the particle's charge fluctuates with a small amplitude of d$Q$ of an alternating sign.
\textcolor{black}{For example, most of the 200~$\upmu$m particles exceeded their equilibrium charge after three impacts.
Therefore, during the fourth impact, 63\% of the particles have a negative impact charge.}
In total, about 72\% of all recorded impacts lead to an impact charge of less than 0.05~pC.
That value may serve as a threshold to identify the impacts when the particle is already saturated.
Out of the impacts above this threshold, only 14\% of the small and 28\% of the large particles charge negatively.
These cases might be caused by contamination of the particle surface or the plates due to the wear of earlier experiments.
Since wear also appears at the duct in the powder flow experiment, the data in Fig.~\ref{fig:ress} of both polarities are valuable for the single-particle charging model.

The charge the particles hold ($Q_i$) depends on their size.
For most of the 100~$\upmu$m particles, $Q_\mathrm{i}$ varies between -0.3~pC and~0.8~pC and for the 200~$\upmu$m particles between -0.7~pC and~1.6~pC.
That means the charge roughly doubles, which is less than expected since the surface areas of both particle sizes differ by a factor of four.
Thus, the particle charges scale non-linearly with the surface area, which might be due to the inhomogeneous charge distribution on their non-conductive surfaces.
On the other hand, the impact charge lies between -0.2~pC and 0.4~pC, independent of the particle size.

The small $Q_\mathrm{i}$ for high $v_\mathrm{i}$, according to Figs.~\ref{fig:a} and~\ref{fig:d}, explains as follows.
The first impacts of the particles on the lower plate are those with the highest velocity (cf.~Fig.~\ref{fig:post}).
Also, before the first impact, the particles carry only their initial charge, which is usually lower than the charge they accumulate in the following impacts.
Surprisingly, no impacts occur in the region 0.7~pC~$<Q_\mathrm{i}<1.4$~pC and $v_\mathrm{i}<12$~m/s in Fig.~\ref{fig:d}.
A similar but less pronounced region appears in Fig.~\ref{fig:a}.
These low-velocity impacts are the last ones taking place close to the end of each experiment.
By then, most particles had already reached their equilibrium charge between~1.4~pC and~1.6~pC during the previous impacts.
The few particles that did not accumulate their equilibrium charge cannot reach it anymore cannot the impact velocities are too low.

Figures~\ref{fig:c} and~\ref{fig:d} reveal the strong dependence of d$Q$ on $Q_\mathrm{i}$.
The impact charge is low if the charge before the impact is either highly positive or negative.
By far, the highest impact charge occurs if the pre-charge is between 0~pC and 0.2~pC, respectively 0.5~pC.
This has two reasons.
First, a highly charged particle cannot take up a high impact charge since the sum of both would exceed its saturation.
Second, as explained above, the impacts of a low pre-charge are often the first ones that are of a high velocity.
With the impact velocity, the contact area and time between the particle and plate increase, which enhances the charge transfer.

To summarize, the PMMA particles charge on the metal target somewhat irregular, which fortifies using our empirical model over other models based on parameterized correlations.

\subsection{Powder charging simulations and experimental validation}
\label{sec:respow}

\begin{figure}[b]
\begin{center}
\begin{tikzpicture}[thick]
\node [anchor=north east,inner sep=0] at (0,0) {\includegraphics[trim=0mm 0mm 0mm 0mm,clip=true,width=0.95\textwidth]{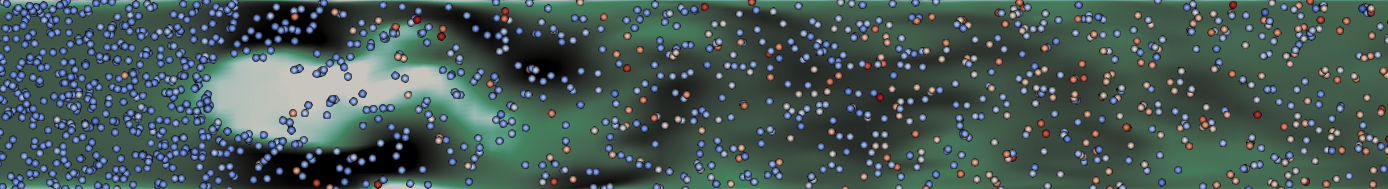}};
\begin{scope}[shift={(-3,2)}]
  \node[anchor=north east,inner sep=0] at (0,0) {\includegraphics[trim=0cm 0cm 0cm 0cm,clip=true,width=5mm,height=12mm]{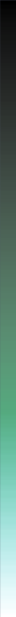}};
	\draw[](-.2,0) node[above]{$u$ / (m/s)};         
	\draw[](0,-.1) node[right]{\small 30};         
	\draw[](0,-1.15) node[right]{\small 0};         
\end{scope}
\begin{scope}[shift={(-1,2)}]
  \node[anchor=north east,inner sep=0] at (0,0) {\includegraphics[trim=0cm 0cm 0cm 0cm,clip=true,width=5mm,height=12mm]{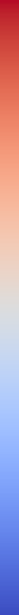}};
	\draw[](-.2,0) node[above]{$Q$ / (pC)};         
	\draw[](0,-.1) node[right]{\small 1.1};         
	\draw[](0,-1.15) node[right]{\small -0.26};         
\end{scope}
\draw[] (-12.86,-2.14) -- (-12.86,-.04);
\draw[fill=gray] (-12.86,-1.1) circle [radius=.39];
\draw [thin] (-12.86,-2.3) -- (-12.86,-2.6);
\draw [<->,>=latex,thin] (-15.6,-2.45) -- (-12.86,-2.45) node[below,midway,align=right] {60 mm};
\draw [<->,>=latex,thin] (-12.86,-2.45) -- (0,-2.45) node[below,midway,align=right] {280 mm};
\draw [->,ultra thick] (-15.6,.7) -- (-12,.7) node[above,midway,align=right] {flow direction};
\end{tikzpicture}
\end{center}
\caption{View from the top on the simulated instantaneous particle and gas flow field.
The centreline velocity of the air at the inlet is 14.7~m/s and the particle mass flow rate is 0.36~kg/h.
The snapshot depicts the last 60~mm of the PMMA duct before the obstacle and 280~mm of the metallic duct thereafter.
The cross-section of the flow field lies in the center of the duct in vertical direction.
The particles are enlarged and only the particles above this plane are shown.
Most particles charge when colliding with the obstacle and in the following metallic duct. 
}
\label{fig:3d}
\end{figure}
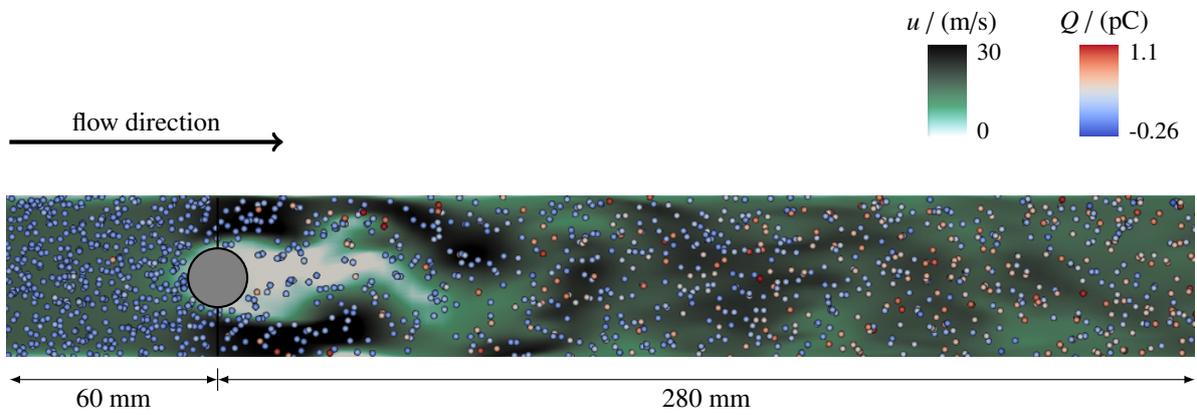

As detailed in Sec.~\ref{sec:singlem}, we used the data of Fig.~\ref{fig:ress} to set up the single-particle charging model for the simulations.
Figure~\ref{fig:3d} visualizes a snapshot of a simulated flow field.
The plotted section, namely the end of the PMMA duct, the obstacle, and the beginning of the metallic duct, is where the flow is the most irregular.
However, the computational domain covers the complete duct (see Fig.~\ref{fig:rig}).
When bypassing the obstacle, the velocity of the carrier gas increases.
Behind the obstacle forms a low-velocity region and a wake that extends far into the metallic duct.
In the simulations, the particles charge only when contacting metallic parts, but not when contacting the PMMA duct.
Nevertheless, in the snapshot, a few particles that carry charge reside upstream of the obstacle.
These few particles impacted the obstacle before, received charge, and bounced back upstream.
Downstream of the obstacle, more particles carry charge, and even more particles obtain charge when flying through the metallic duct toward the outlet.

\begin{figure}[b]
\centering
\subfigure[Experiment, $d_\mathrm{p}=200~\upmu$m]{\includegraphics[trim=0mm 0mm 0mm 0mm,clip=true,width=0.47\textwidth]{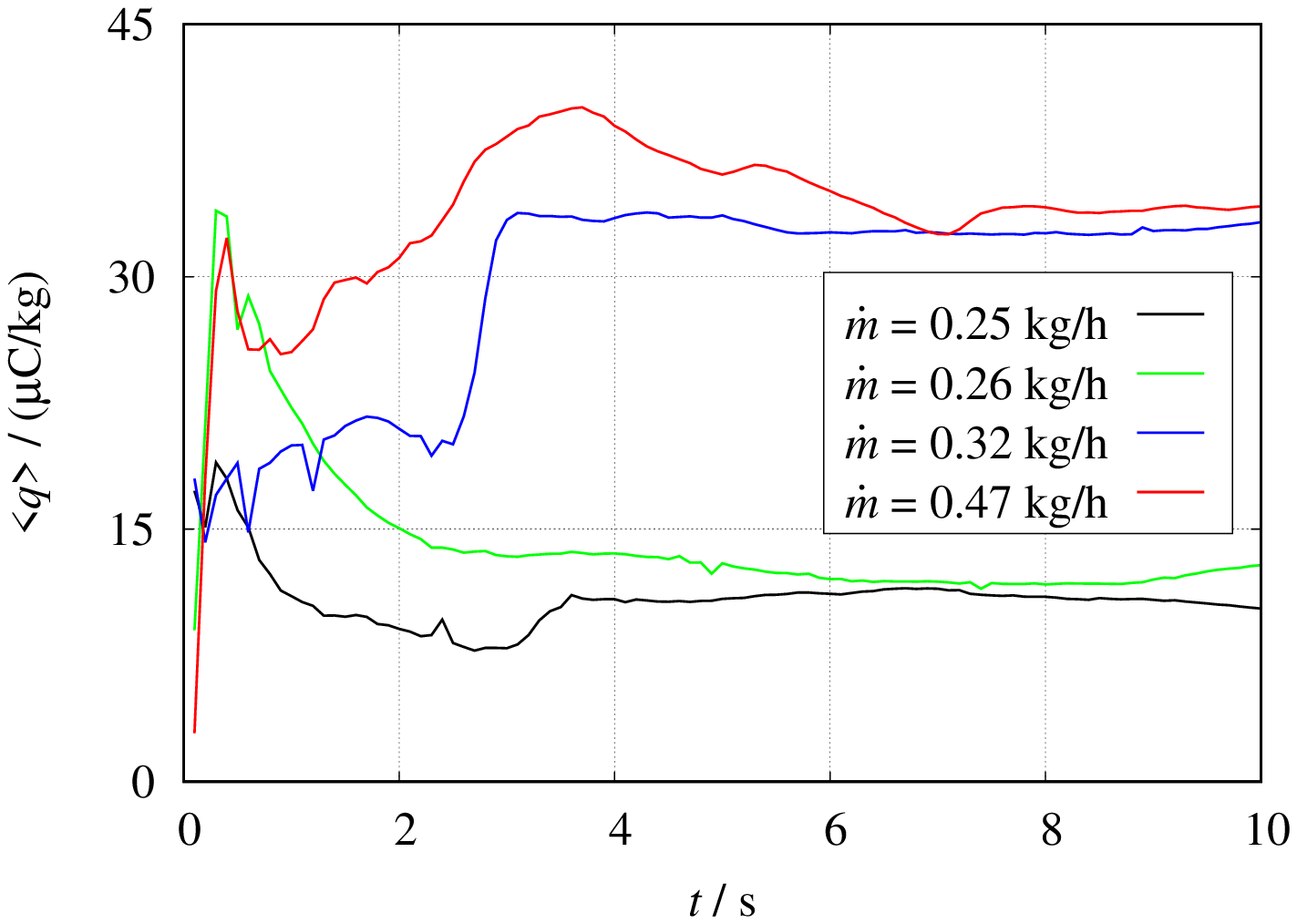}\label{fig:convergencea}}\quad
\subfigure[Simulation]{\includegraphics[trim=0mm 0mm 0mm 0mm,clip=true,width=0.47\textwidth]{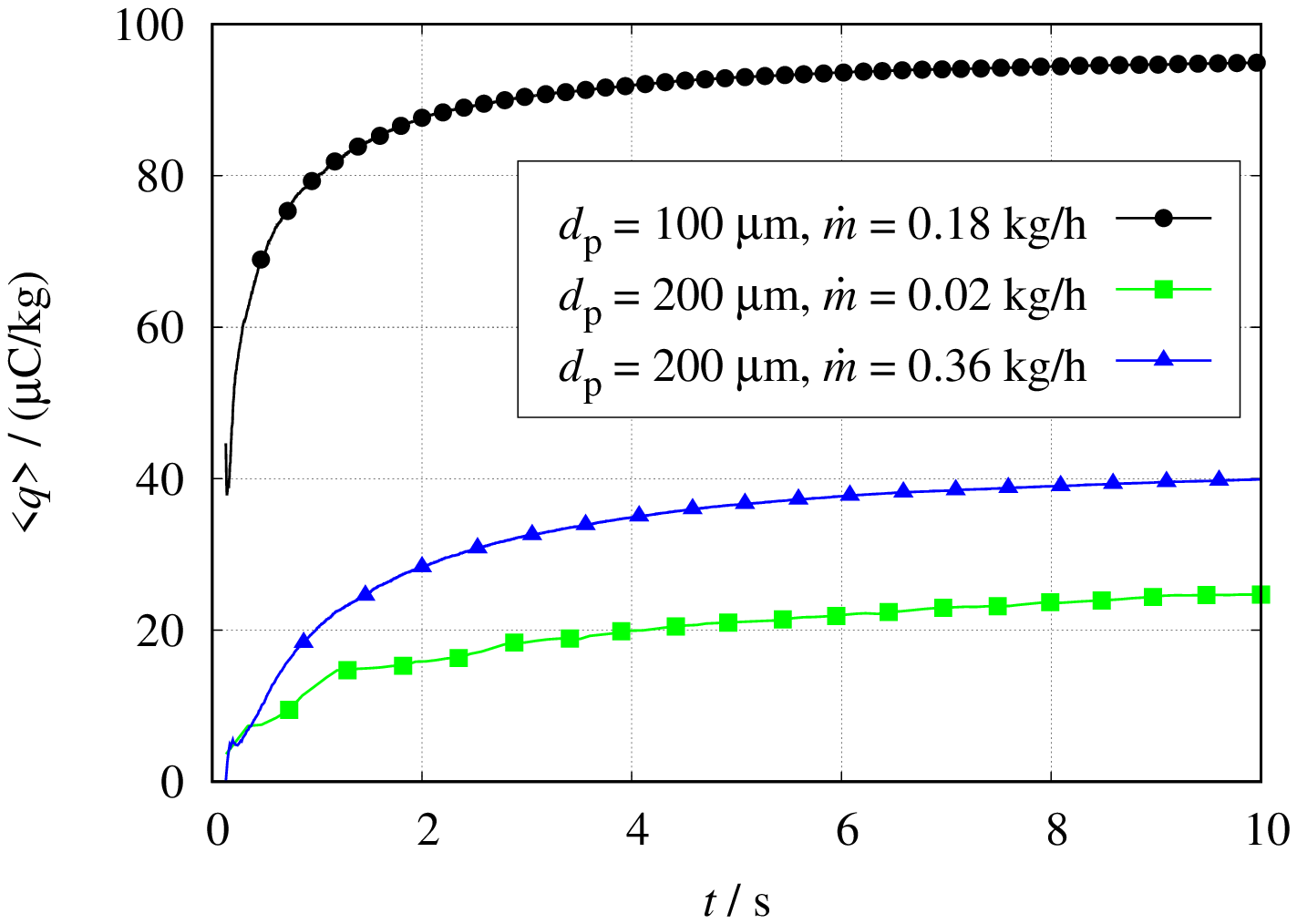}\label{fig:convergenceb}}
\caption{(a) Measured and (b) simulated specific charge of the powder that left the duct versus the conveying time.
The given charge at time $t$ represents the total charge of the particles that left the duct at the outlet between the start of conveying and $t$.}
\label{fig:convergence}
\end{figure}

Since turbulent flows fluctuate chaotic, we compare the simulations with experiments based on statistically converged quantities.
The measured quantity in the powder flow experiment is the total charge of the powder that a filter captures inside the Faraday after leaving the duct.
After finishing the experiment, we weighed the mass of the captured powder to determine its specific charge.
Figure~\ref{fig:convergencea} depicts the evolution of the specific charge inside the Faraday for five experiments, all using 200~$\upmu$m particles.
According to these curves, the charge of the particles collected in the Faraday fluctuates in the beginning but converges latest after about six seconds.
Also, the simulated curves Fig.~\ref{fig:convergenceb} converge and change only slightly after about six seconds.
The symbols assigned to the curves indicate the number of particles leaving the duct.
For a mass flow rate of 0.18~kg/h and a particle size of 100~$\upmu$m, 15\,000 particles form the curve in-between two symbols.
For 0.36~kg/h and 200~$\upmu$m, 10\,000 particles passed in-between two symbols, and for 0.02~kg/h and 200~$\upmu$m only 500 particles passed.
The flow with the highest number of particles per time converges the fastest and the one with the lowest number the slowest.
In other words, the qualitative difference between the experimental and simulated curves is due to technical details of the upstart of the particle feeder and initial conditions to the simulations.
Therefore, we simulate and measure the evolution of the specific charge for ten seconds.
Through this procedure, we compare the measurements with simulations based on statistically converged data.

\begin{figure}[tb]
\centering
\subfigure[$d_\mathrm{p}=100~\upmu$m]{\includegraphics[trim=0mm 0mm 0mm 0mm,clip=true,width=0.47\textwidth]{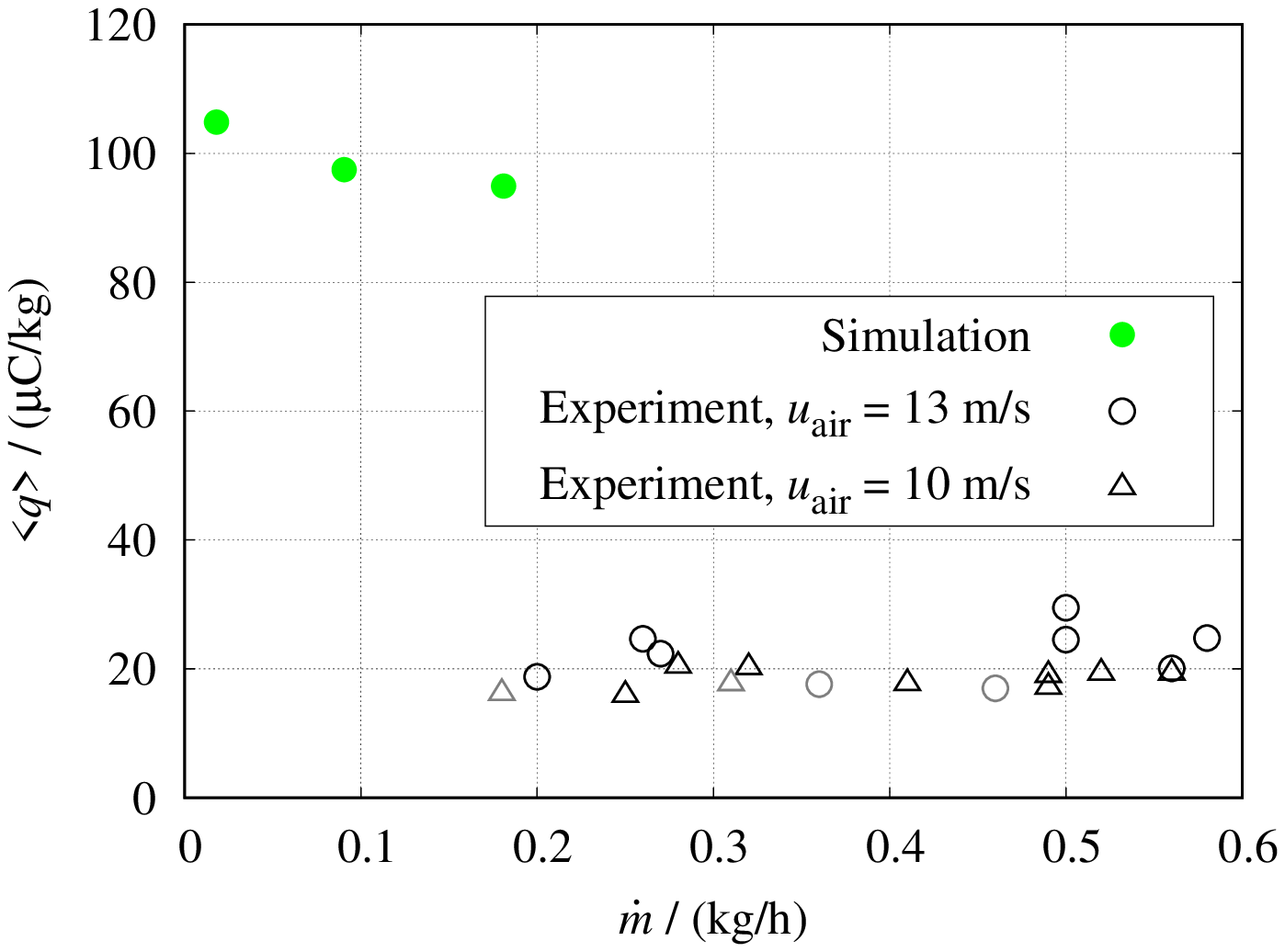}\label{fig:validation100}}\quad
\subfigure[$d_\mathrm{p}=200~\upmu$m]{\includegraphics[trim=0mm 0mm 0mm 0mm,clip=true,width=0.47\textwidth]{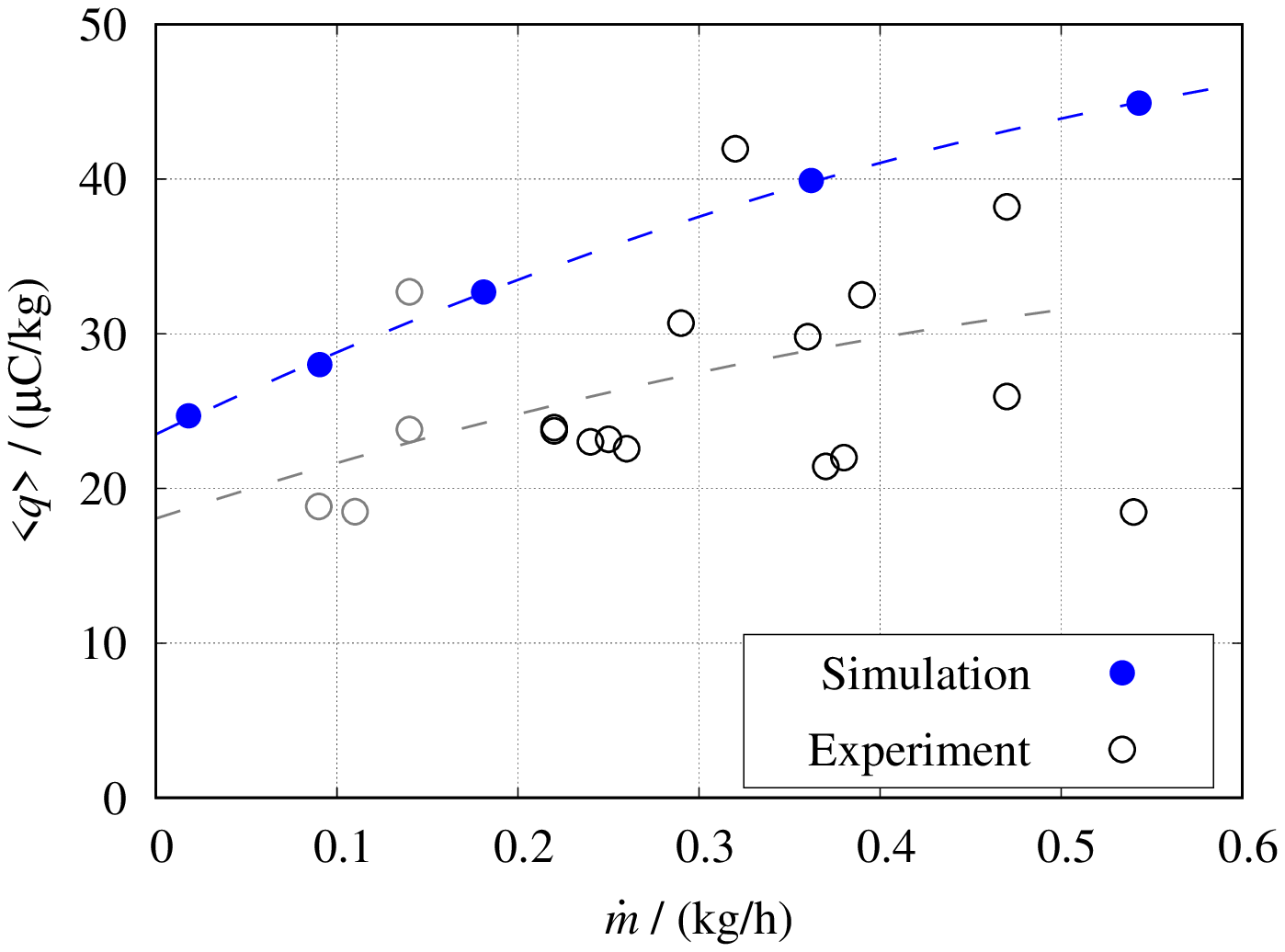}\label{fig:validation200}}
\caption{Comparison of the specific powder charge versus powder mass flow rate computed with the simulations and measured in the experiments.
The grey symbols indicate experiments of a lower quality, the dashed lines second-order fitted polynomials.}
\label{fig:validation}
\end{figure}
At first glance, in Fig.~\ref{fig:validation}, the simulations agree well with the experiments for the 200~$\upmu$m particles.
For the 100~$\upmu$m particles, the differences are more significant.
The depicted charge is collected only in the metallic part of the duct.
That means we subtracted from the total measured charge the average charge accumulated in the PMMA duct, which we determined in separate measurements.
Accordingly, we subtracted the initial from the total simulated charge.

We measured the 100~$\upmu$m particles during two runs.
\textcolor{black}{During the first run, the ambient temperature was 18~$^{\circ}$C and the relative humidity was 55\%, during the second one 20~$^{\circ}$C and 52\%.}
The centreline velocity of the air was 13~m/s, which is close to the simulated one.
\textcolor{black}{According to the earlier measured velocity profile~\citep{Gro20c}, this velocity corresponds to an airflow rate of about~1.32~m$^3$/min.}
We used a finer filter for the second test run that caused a higher pressure drop and resulted in a slightly lower air velocity.
\textcolor{black}{The experiments with the 200~$\upmu$m particles took place in an ambiance of~19~$^{\circ}$C to~20~$^{\circ}$C and 52\% to 60\%; the centreline velocity of the air was 13~m/s.}

\textcolor{black}{
These conveying velocities are at least twice as high as the particles' deposition velocities.
In the earlier experiments, a conveying velocity above 6.5~m/s prevented 100~$\upmu$m particles from sticking to the duct, and a conveying velocity above 5.2~m/s prevented 200~$\upmu$m particles from sticking.
Since most of the duct is transparent, we could optically confirm that all particles reached the Faraday and no deposits formed.}

In Fig.~\ref{fig:validation}, the black symbols indicate high-quality experiments, whereas the grey symbols mark problematic tests.
In the problematic tests, the voltage curves recorded from the Faraday were difficult to interpret because of unusual peaks or high fluctuations.
Nearly all these tests regarded low powder mass flow rates, where the charge was averaged over relatively few particles.
Furthermore, the low mass flow rates are the lower end of the operating conditions of our particle feeder.
Therefore, the powder supply probably fluctuated more than for higher mass flow rates.

This limitation of the experiment partly explains why we did not simulate and measure the 100~$\upmu$m particles for the same range of mass flow rates (Fig.~\ref{fig:validation100}).
The other reason is that the computational expense of the simulations scales with the number of tracked particles.
Thus, we refrained from simulating higher mass flow rates than those presented.
It is noted that the same number of particles corresponds to an eight times higher mass flow rate of 200~$\upmu$m particles.
Therefore, we could simulate and measure a comparable range of mass flow rates for this particle size.

\textcolor{black}{
The observed deviations point to possibly lacking physics in our simulation model in the case of many particles.
A locally high concentration of charged particles leads to a high electric field.
According to the charge relaxation theory~\citep{Mat95b}, a surrounding electric field reduces the charge a particle can hold.
In the experiment, relaxation might reduce the charge of those flows containing the most particles.
That means all experiments with 100~$\upmu$m particles and those with 200~$\upmu$m particles and high mass flow rates, for which the charge in Fig.~\ref{fig:validation200} drops.
But charge relaxation is not implemented in our numerical tool, which can be one reason for over-predicting the charge.}

Another reason for the difference between experiments and simulations can be the particle charging model.
To evaluate the model, we plotted 15\,000 impact conditions from the simulations in Figs.~\ref{fig:a} and~\ref{fig:d} together with the conditions of the single-particle experiment.
In general, the simulation conditions overlap the ones recorded in the experiment.
In other words, our simulation model predicts charge exchange between a particle and a surface-based on similar experimental conditions.

Nevertheless, a representative statistical base crucially enhances the accuracy of the model predictions.
Most experimental impacts in Figs.~\ref{fig:a} and~\ref{fig:d} lie out of the grey area.
Thus, the model does not use these data.
Even though unused data points do not impair the model, a higher density of experimental impacts within the range of simulation conditions is desirable.
Besides reducing the number of experiments, that would improve the statistical base of the model.
The above discussion points directly toward the future development of single-particle charging experiments to support model development.
First, impact data for lower impact velocities would increase the statistical base and, therefore, the model's accuracy.

Especially, according to Fig.~\ref{fig:a}, many 100~$\upmu$m particles contact the walls slower than in the experiment.
The experiment yields no impact velocities of less than 2~m/s.
Therefore, for these velocities, the single-particle charging model performs poorly.
This is another reason for the deviations between experiments and simulations for this particle size.
Nonetheless, the data for the 200~$\upmu$m particles agrees well.
Once again, the agreement in Fig.~\ref{fig:validation200} of the absolute charge and its slope are remarkable, considering we tuned no parameter.
Given the success of validating the simulations with the 200~$\upmu$m particles and the flaws of the simulations with the 100~$\upmu$m particles, the following focuses on the simulations with the larger particles.

\begin{figure}[tb]
\centering
\subfigure[]{\includegraphics[trim=0mm 0mm 0mm 0mm,clip=true,width=0.47\textwidth]{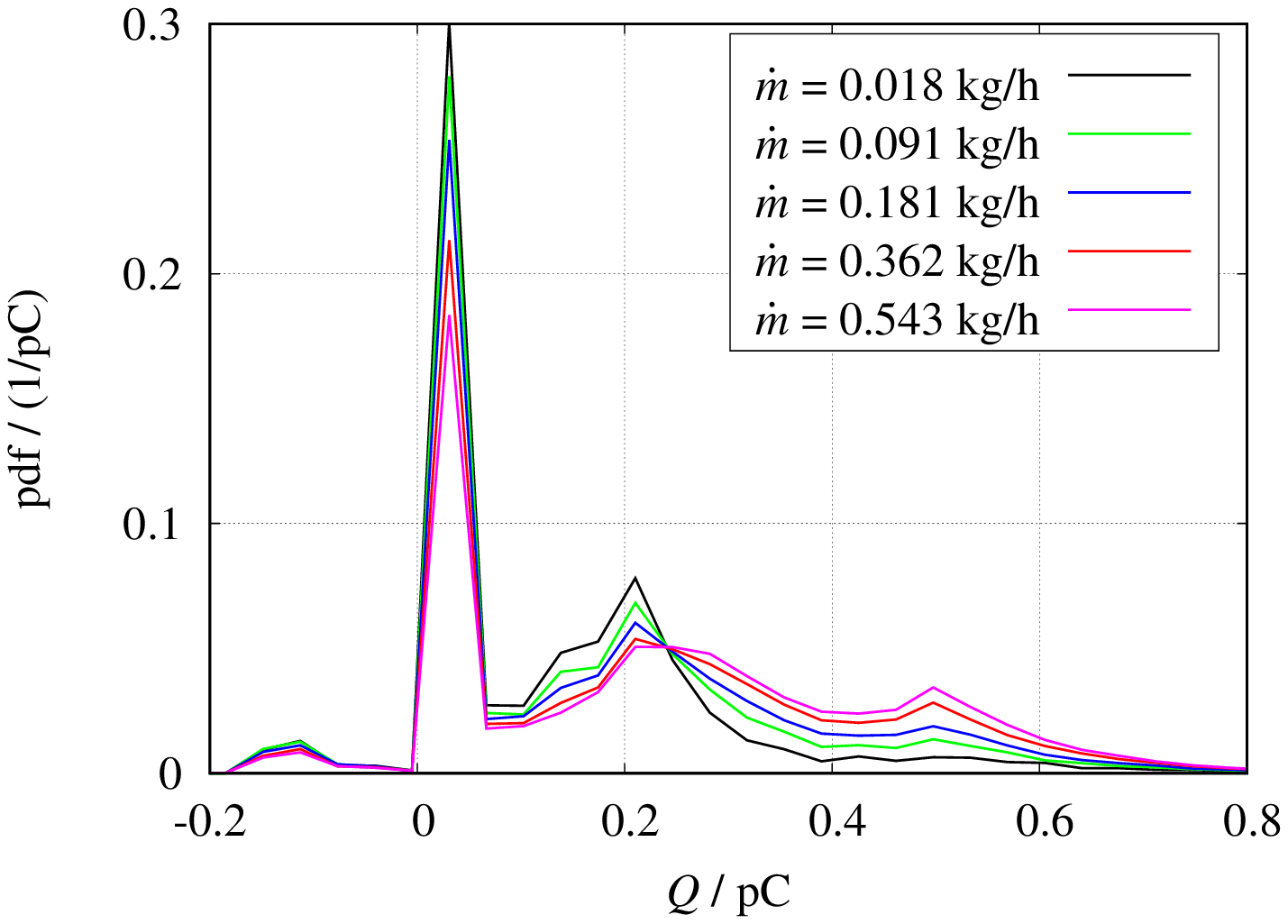}\label{fig:pdf}}\quad
\subfigure[]{\includegraphics[trim=0mm 0mm 0mm 0mm,clip=true,width=0.47\textwidth]{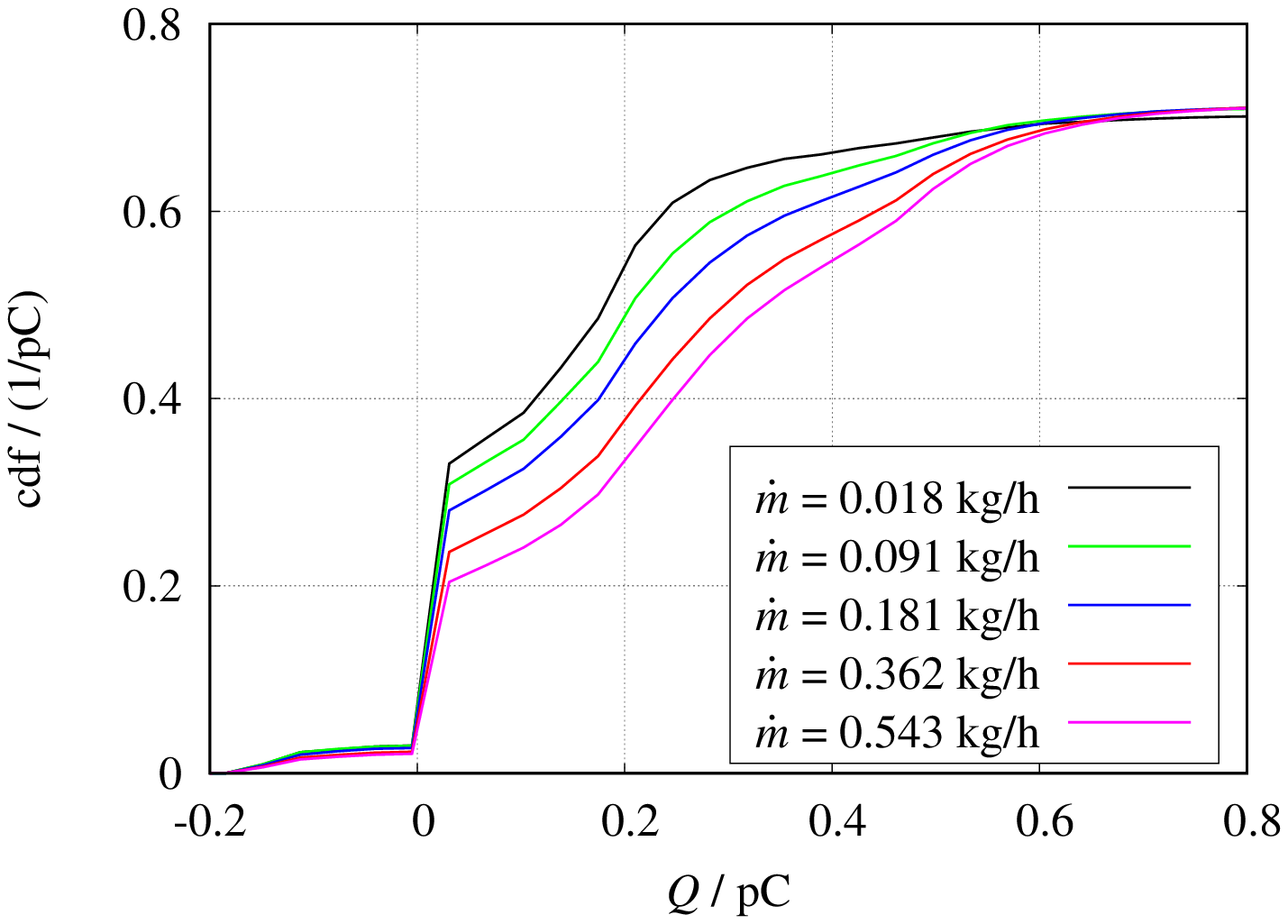}\label{fig:cdf}}
\caption{Bipolar charge distributions predicted by the simulations with 200~$\upmu$m particles.
(a) Probability density function (pdf) and (b) cumulative distribution function (cdf) of the particle charge after leaving the duct.}
\label{fig:pdfcharge}
\end{figure}
The detailed analysis of the simulations shows that the powder charges bipolar.
Figure~\ref{fig:pdfcharge} depicts the charge distribution of the simulated 200~$\upmu$m powder batch depending on the mass flow rate, once plotted as (\ref{fig:pdf})~probability density function~(pdf) and once as (\ref{fig:cdf})~cumulative distribution function~(cdf).
According to these figures, most particles charge positively in the metallic duct.
But for all flow rates, some particles form the negative tail of the distributions.
In other words, all particles enter the metallic duct carrying a positive initial charge, but some change their sign during downstream contacts.
That dynamics stems from the data in the fourth quadrant of Fig.~\ref{fig:d}, namely the impacts of positively charged particles that received negative impact charge.
Thus, contrary to conventional particle charging models, our model predicts bipolar charging.

The positive sides of the charge distributions of all flow rates peak at 0.025~pC, which is the initial charge assigned to the 200~$\upmu$m particles.
Hence, those particles retained their charge while traveling through the metallic duct, probably because they neither contacted the walls nor the obstacle.
Between 0.1~pC and 0.25~pC is the region of the mildly charged particles.
The lower the mass flow rate, the higher the number of mildly charged particles.
The turnover point lies at approximately 0.25~pC.
Beyond that point, the tendency reverses; the higher the mass flow rate, the more highly charged particles.

\begin{figure}[tb]
\centering
\subfigure[]{\includegraphics[trim=0mm 0mm 0mm 0mm,clip=true,width=0.47\textwidth]{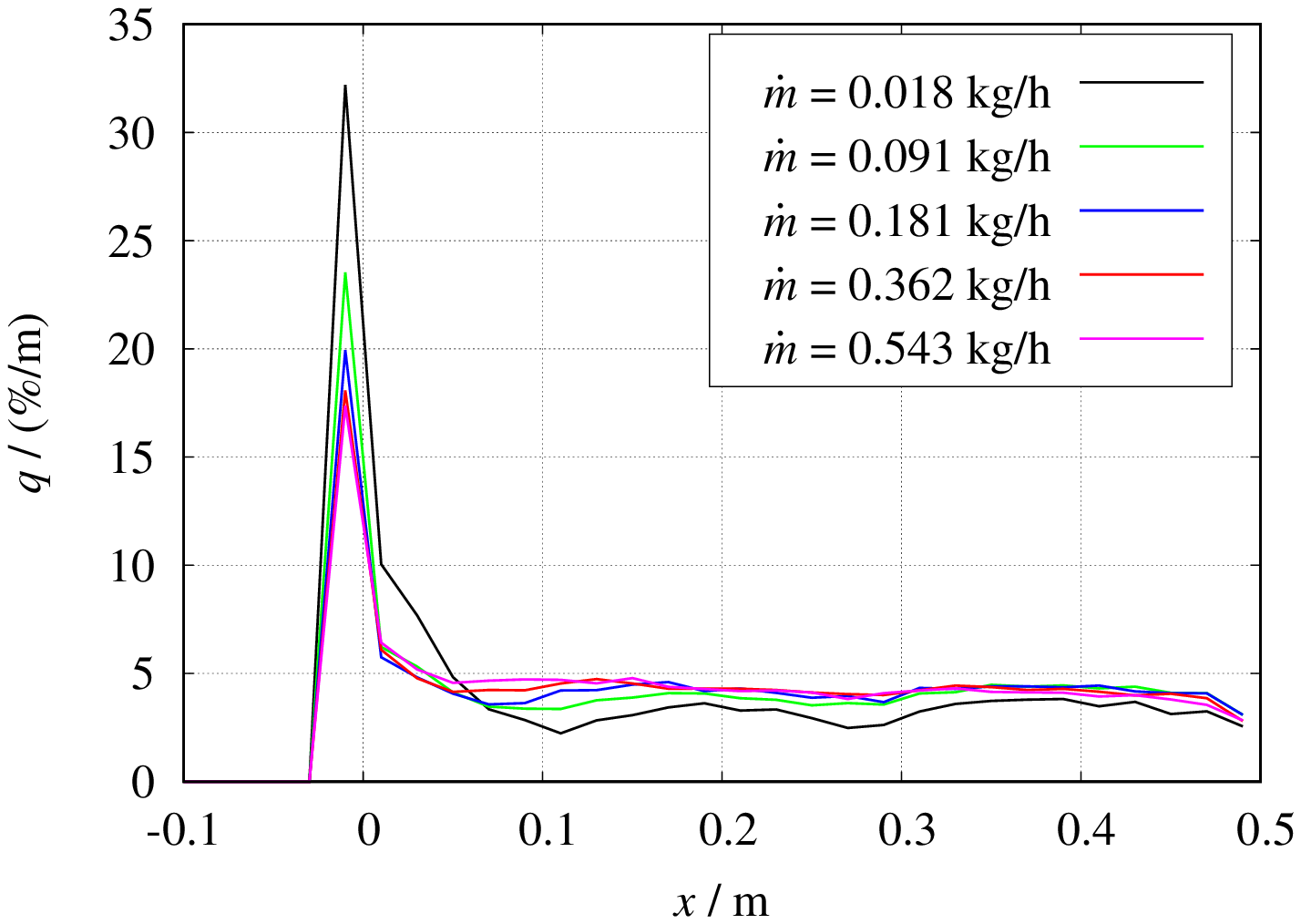}\label{fig:qn}}\quad
\subfigure[]{\includegraphics[trim=0mm 0mm 0mm 0mm,clip=true,width=0.47\textwidth]{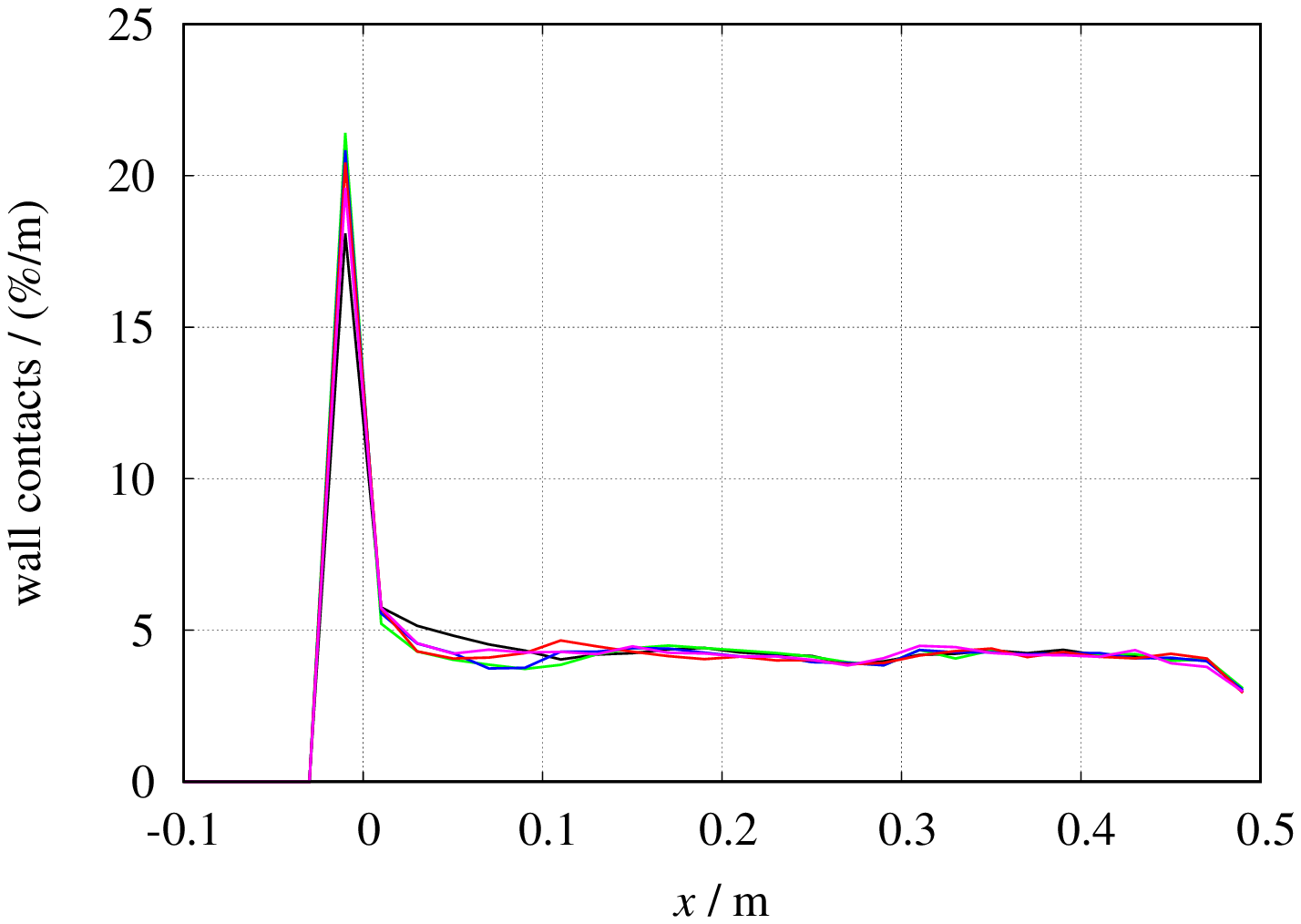}\label{fig:hnpn}}\\
\subfigure[]{\includegraphics[trim=0mm 0mm 0mm 0mm,clip=true,width=0.47\textwidth]{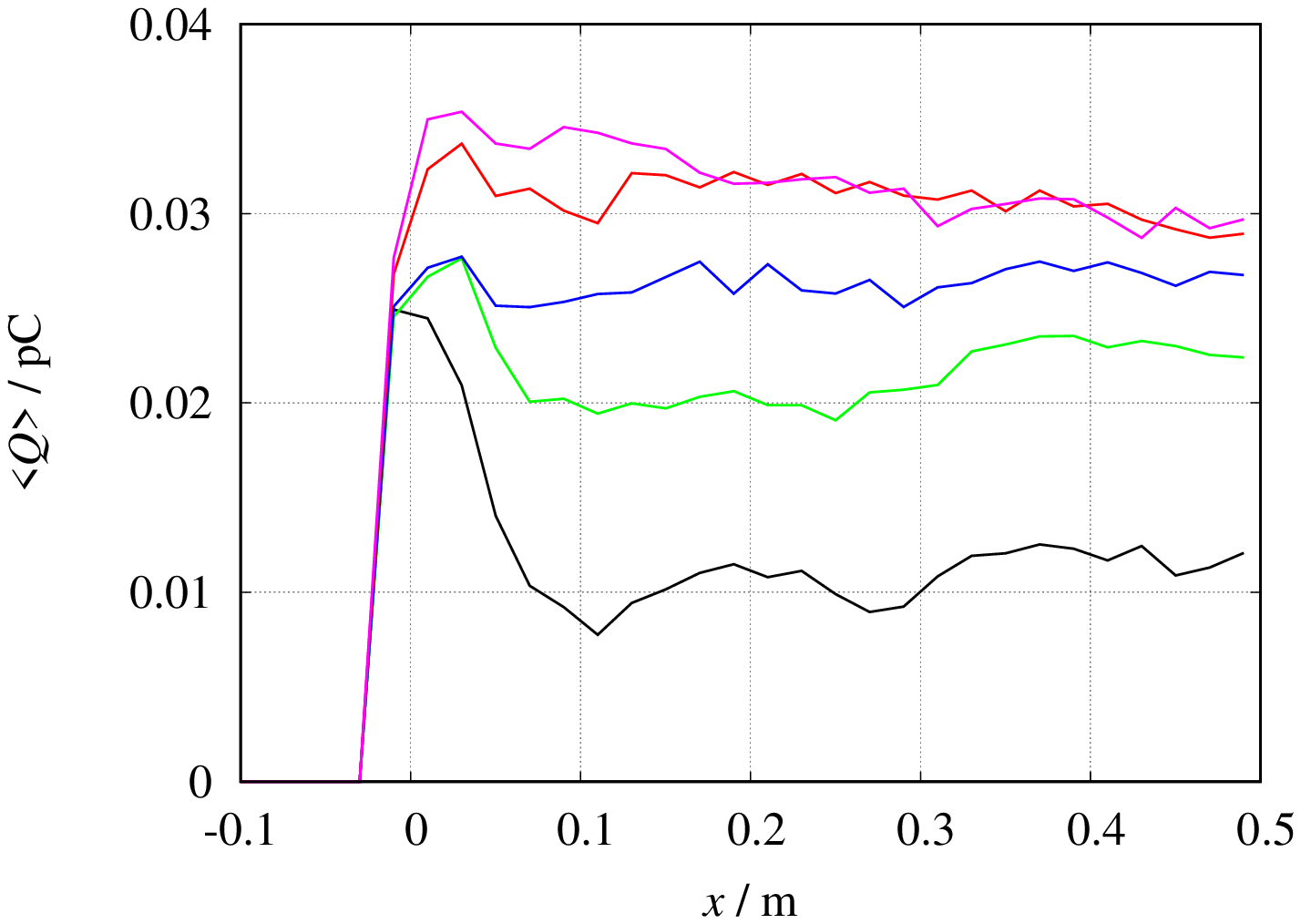}\label{fig:qav}}\quad
\subfigure[]{\includegraphics[trim=0mm 0mm 0mm 0mm,clip=true,width=0.47\textwidth]{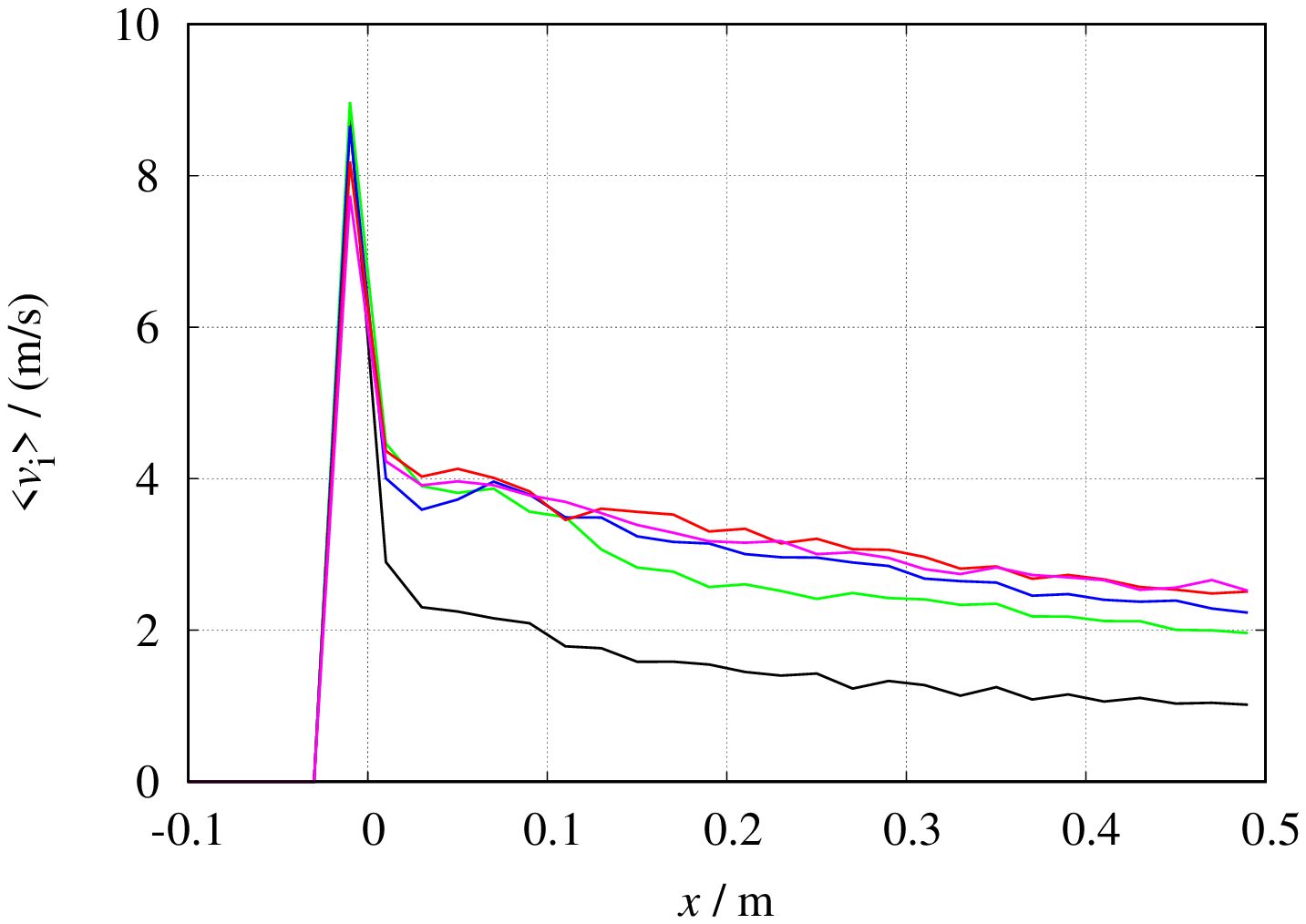}\label{fig:uav}}
\caption{Results of simulations with 200~$\upmu$m particles. The origin of the $x$-axis is at the beginning of the metallic duct.
The plots give, depending on the streamwise location in the duct, 
(a) charge generation per meter as a fraction of the total powder charge,
(b) number of particle-wall contacts per meter as a fraction of the total number of contacts,
(c) average charge accumulation per contact, and
(d) average impact velocity per contact.}
\label{fig:dq}
\end{figure}
Further, we analyzed the location of charge generation based on the simulations with the 200~$\upmu$m particles.
Figure~\ref{fig:qn} visualizes the charge taken up by the powder per unit distance traveling through the metallic duct.
Therein, the charge amount is given as a fraction of the final charge of each powder batch.
Since the charging of the powder in the PMMA duct is not part of our simulations, all plots in Fig.~\ref{fig:dq} focus on the metallic parts.
The origin of the spatial coordinate, $x=0$, refers to the beginning of the metallic duct, which is also the obstacle's center axis.
According to Fig.~\ref{fig:qn}, all flows generate by far the highest charge in this region. 
The obstacle extends partly into the PMMA duct (cf.~Figure~\ref{fig:obstacle}), which explains why the powder charges already before reaching the beginning of the duct ($x<0$).
In short, the particles mostly charge while passing the obstacle and less in the downstream part of the metallic duct.

Furthermore, Fig.~\ref{fig:qn} reveals that particles charge less at the obstacle and more in the duct's downstream section if the mass flow rate is high.
This trend appears either because the mass flow rate affects the location of the particle-wall contacts or the amount of transferred charge during each contact.
With this in mind, Fig.~\ref{fig:hnpn} depicts the number of particle-wall contacts per unit distance, normalized to the total number of contacts.
When comparing the curves, the mass flow rate influences only marginally where the particles contact the duct.
Thus, we next analyze the average charge exchange between a particle and a wall or the obstacle upon one contact.
As inferred from Fig.~\ref{fig:qav}, the average impact charge at the windward side of the obstacle is similar for all flow rates.
However, downstream of the obstacle the impact charge of the lowest mass flow drops fast and is higher for the cases with a higher mass flow rate.
According to Fig.~\ref{fig:uav}, the high impact charge in the downstream part results from the high impact velocity for high mass flow rates.
Thus, the faster charging with high particle flow rates results from the increased wall-normal velocities of the particles.

This is precisely the information we need to prevent explosions effectively.
Knowing the conveying system’s locations most prone to charge generation puts forward countermeasures.
A possible countermeasure to reduce the charge of the conveying system is bonding the cylindrical obstacle.
To reduce the charging of the powder, one can change the shape of the obstacle or remove it.
This analysis takes advantage of developing models through single-particle and powder flow experiments and the details provided by simulations.

\section{Conclusions}

Our three-part approach to explore powder flow electrification, including single-particle experiments, CFD simulations, and powder flow experiments, holds several advantages.
\textcolor{black}{Motivated by the scatter of the single-particle experiment's impact data, we proposed an empirical charging model.}
Using this model, the simulated charge of 200~$\upmu$m sized PMMA particles agreed well with the measurements of the powder flow experiments.
Contrary to previous attempts, our model does not require tuning any parameters.
Further, our simulations predicted bipolar charge distributions.
Most particles charged when hitting a cylindrical obstacle installed inside the duct, \textcolor{black}{highlighting the influence of built-in equipment on electrification}.
Generally speaking, with this model, we reached a realistic and detailed picture of how, where, and why powder electrifies during pneumatic conveying.
The possibility to extract such details holds great promise for the safety analysis of powder processing.
But equally important, our approach pinpoints toward the research needed for further improvement.
\textcolor{black}{First, including charge relaxation in the simulations may improve the predictions for powder flows that generate a locally strong electric field, such as for the 100~$\upmu$m particles in our study}.
Second, a better overlap of the impact conditions of the single-particle experiment and the simulations will instantly increase the accuracy of the particle charging model.
Third, although our model reflects the most important factors influencing the impact charge, namely pre-charge and impact velocity, including more parameters enhances the model's generality.
Thus, the challenge for obtaining a predictive simulation tool is bringing the operating conditions of the three parts closer to each other.

\section*{Acknowledgments}

This project has received funding from the European Research Council~(ERC) under the European Union’s Horizon 2020 research and innovation programme~(grant agreement No.~947606 PowFEct).
The first author gratefully acknowledges the support of the research visit at Soka University through the PTB guest researcher fellowship.

\bibliography{\string~/essentials/publications}

\end{document}